\documentclass[10pt, twocolumn]{IEEEtran}

\IEEEoverridecommandlockouts

\usepackage{cite}
\usepackage{graphicx}
\usepackage[cmex10]{amsmath}
\usepackage{amssymb}
\usepackage{array}
\usepackage{psfrag}
\usepackage{arydshln}
\usepackage{pifont}
\usepackage{cancel}
\usepackage{wasysym}
\usepackage{color}

\usepackage{multirow}
\usepackage[pdftex,hypertexnames=false]{hyperref}

\usepackage{footnote}
\makesavenoteenv{tabular}

\newtheorem{lemma}{Lemma}

\newtheorem{proposition}{Proposition}

\newcommand{\mb}{\mathbf}

\newcommand{\uh}{{\mb{h}}}
\newcommand{\uha}{{\uh_1}}

\newcommand{\uhad}{{\uh_{\text{d},1}}}

\newcommand{\uhar}{{\uh_{\text{r},1}}}

\newcommand{\uhda}{{\mb{h}_{\text{d},1}}}

\newcommand{\uhra}{{\mb{h}_{\text{r},1}}}

\newcommand{\ur}{{\mb{r}}}

\newcommand{\up}{{\mb{p}}}

\newcommand{\uQ}{{\mb{Q}}}

\newcommand{\uf}{{\mb{f}}}

\newcommand{\uH}{{\mb{H}}}
\newcommand{\uHH}{{\mb{H}}^{\cal{H}}}

\newcommand{\uHb}{{\mb{H}_2}}

\newcommand{\uHd}{{\uH_{\text{d}}}}

\newcommand{\uHdb}{{\uH_{\text{d},2}}}

\newcommand{\uHdn}{{\uH_{\text{d,n}}}}

\newcommand{\uHr}{{\uH_{\text{r}}}}

\newcommand{\uHrb}{{\uH_{\text{r},2}}}

\newcommand{\uHrn}{{\uH_{\text{r,n}}}}

\newcommand{\NT}{{N_{\text{T}}}}

\newcommand{\NR}{{N_{\text{R}}}}
\newcommand{\uP}{{\mb{P}}}
\newcommand{\uR}{{\mb{R}}}

\newcommand{\uRT}{{\mathbf{R}_{\text{T}}}}

\newcommand{\uRTK}{{\mathbf{R}_{\text{T,}K}}}

\newcommand{\uv}{{\mb{v}}}

\newcommand{\ux}{{\mb{x}}}

\newcommand{\mzero}{\mb{0}}

\newcommand{\uF}{{\mb{F}}}

\newcommand{\mbI}{\mb{I}}

\newcommand{\RTinvaa}{\left[\mathbf{R}_{\text{T},K}^{-1}\right]_{1,1}}

\newcommand{\uqsimple}{\mathbf{q}}

\title{MIMO Zero-Forcing Performance Evaluation \\Using the Holonomic Gradient Method}

\author{
Constantin Siriteanu\thanks{C.~Siriteanu is with the Department of Information Systems Engineering, Graduate School of Information Science and Technology, Osaka University.}, 
Akimichi Takemura\thanks{A.~Takemura is with the Department of Mathematical Informatics, Graduate School of Information Science and Technology, University of Tokyo, Japan, and Japan Science and Technology Agency, CREST.},
Satoshi Kuriki\thanks{S.~Kuriki is with the Institute of Statistical Mathematics, Tachikawa, Tokyo, Japan.}, 
Hyundong Shin\thanks{H.~Shin is with the Department of Electronics and Radio Engineering, Kyung Hee University, South Korea.},
Christoph Koutschan\thanks{C. Koutschan is with the Johann Radon Institute for Computational and Applied Mathematics, Austrian Academy of Sciences, Linz, Austria.}}

\begin{document}

\maketitle

\begin{abstract}
For multiple-input multiple-output (MIMO) spatial-multiplexing transmission, zero-forcing detection (ZF) is appealing because of its low complexity.
Our recent MIMO ZF performance analysis for Rician--Rayleigh fading, which is relevant in heterogeneous networks, has yielded for the ZF outage probability and ergodic capacity infinite-series expressions. Because they arose from expanding the confluent hypergeometric function $ {_1\! F_1} (\cdot, \cdot, \sigma) $ around 0, they do not converge numerically at realistically-high Rician $ K $-factor values. 
Therefore, herein, we seek to take advantage of the fact that $ {_1\! F_1} (\cdot, \cdot, \sigma) $ satisfies a differential equation, i.e., it is a \textit{holonomic} function. 
Holonomic functions can be computed by the \textit{holonomic gradient method} (HGM), i.e., by numerically solving the satisfied differential equation. Thus, we first reveal that the moment generating function (m.g.f.) and probability density function (p.d.f.) of the ZF signal-to-noise ratio (SNR) are holonomic. Then, from the differential equation for $ {_1\! F_1} (\cdot, \cdot, \sigma) $, we deduce those satisfied by the SNR m.g.f. and p.d.f., and demonstrate that the HGM helps compute the p.d.f. accurately at practically-relevant values of $ K $. Finally, numerical integration of the SNR p.d.f. produced by HGM yields accurate ZF outage probability and ergodic capacity results.
\end{abstract}


\begin{IEEEkeywords}
Holonomic gradient method, hypergeometric function, MIMO, Rayleigh--Rician fading, zero-forcing.
\end{IEEEkeywords}

\section{Introduction}
\label{section_introduction}

\subsection{Background}
\label{section_background}

The performance of multiple-input multiple-output (MIMO) wireless communications systems has been attracting substantial interest\cite{paulraj_pieee_04}\cite{simon_alouini_book_00}\cite{gesbert_spm_07}\cite{mckay_tcomm_09}\cite{hoydis_jsac_13}. 
Typically, its evaluation proceeds from expressions of performance measures --- e.g., outage probability, ergodic capacity --- derived based on statistical assumptions about the channel matrix\cite{paulraj_pieee_04}\cite{simon_alouini_book_00}\cite{mckay_tcomm_09}\cite{louie_tcom_09}\cite{gore_cl_02}\cite{kiessling_spawc_03}\cite{Matthaiou_jsac_13}\cite{basnayaka_twc_13}\cite{siriteanu_tvt_11}\cite{siriteanu_twc_13}\cite{siriteanu_ausctw_14}.

For tractability, MIMO analyses have often assumed zero-mean channel matrix, i.e., Rayleigh fading\cite{simon_alouini_book_00}. However, state-of-the-art channel measurements and models, e.g., WINNER II\cite{winner_d_1_1_2_v_1_2}, have revealed nonzero-mean channel, i.e., Rician fading\cite{simon_alouini_book_00}.
MIMO performance analysis for Rician fading is much more complicated than for Rayleigh fading.
For linear interference-mitigation approaches such as zero-forcing detection (ZF) and minimum mean-square error detection (MMSE)\cite{bai_choi_book_12}, the performance analysis of MIMO spatial-multiplexing transmission has been found tractable for Rician--Rayleigh mixtures\cite{mckay_tcomm_09}\cite{siriteanu_twc_13} (but not for general Rician fading).

ZF has been considered for WiMAX and LTE\cite{li_cm_10} and has recently been studied as relevant for distributed and large MIMO systems\cite{hoydis_jsac_13}\cite{jung_jcn_13}\cite{Matthaiou_jsac_13}\cite{ngo_tcom_13}\cite{basnayaka_twc_13}, mainly due to its low complexity. 
Nevertheless, although simple, for well-conditioned MIMO channel matrix, ZF approaches the performance of maximum-likelihood and minimum-error-rate (i.e., optimal) detection\cite{artes_tsp_03}\cite{gesbert_tsp_03}\cite{ma_tit_08}\cite{maurer_tsp_09}. 
Thus, the analysis of ZF performance for Rician fading has remained of interest --- see\cite{siriteanu_tvt_11}\cite{siriteanu_twc_13} and references therein.

\subsection{Previous MIMO ZF Performance Analyses}
\label{section_previous_work}

For perfectly-known uncorrelated Rayleigh fading channel, MIMO ZF performance was first characterized exactly in\cite{winters_tcom_94}, by viewing ZF as the no-noise limit of optimum combining, i.e., MMSE\cite[Remark~1]{louie_tcom_09}. 
Recently, though, it has been shown that performance measures for MMSE do not necessarily converge to those of ZF\cite{li_tit_06}\cite{jiang_tit_11}\cite{mehana_tit_12}.

The results of\cite{winters_tcom_94} were extended to transmit-correlated fading in\cite{gore_cl_02}\cite{kiessling_spawc_03} based on the fact that, given the channel matrix $ \uH $, the signal-to-noise ratio (SNR) for ZF is determined by matrix $ \uHH \uH $, which is central-Wishart-distributed when $ \uH $ has zero mean and receive-correlation.

ZF has been studied for Rician fading, i.e., nonzero-mean $ \uH $, much less than for Rayleigh fading.
This is because the analysis is complicated by the noncentral-Wishart distribution of $ \uHH \uH $.
Some approximation-based results for full-Rician fading have appeared for ZF in\cite{siriteanu_tvt_11} and relevant references therein. 
For MMSE, approximate and exact analyses under Rician--Rayleigh fading mixtures appeared in\cite{louie_tcom_09} and\cite{mckay_tcomm_09}, respectively.

Note that, although such fading mixtures have mainly been considered because they promote analysis tractability, they are nevertheless relevant in macrocells, microcells, and heterogeneous networks --- see\cite{mckay_tcomm_09}\cite{siriteanu_twc_13} and references therein.

Thus, we have also analyzed ZF for such fading mixtures in\cite{siriteanu_twc_13}\cite{siriteanu_ausctw_14}.
For the case when the intended stream undergoes Rician fading whereas the interfering streams undergo Rayleigh fading,\cite{siriteanu_twc_13} derived exact infinite-series expressions for ZF performance measures, and\cite{siriteanu_ausctw_14} proved their theoretical convergence.

The exact ZF analysis procedure we employed in\cite{siriteanu_twc_13} is as follows. 
First, we expressed, in\cite[Eq.~(31)]{siriteanu_twc_13}, the moment generating function (m.g.f.) of the SNR for Rician-fading Stream 1 in terms of the confluent hypergeometric function $  {_1\! F_1} (\cdot, \cdot, \sigma) $\cite[Ch.~13]{NIST_book_10}.
Then, the well-known expansion of $ {_1\! F_1} (\cdot, \cdot, \sigma) $ around $ 0 $  from\cite[Eq.~(13.2.2), p.~322]{NIST_book_10} yielded the infinite series for the SNR m.g.f.~from\cite[Eq.~(37)]{siriteanu_twc_13}.
Upon inverse-Laplace transformation, the latter yielded the infinite-series expression for the SNR probability density function (p.d.f.) from\cite[Eq.~(39)]{siriteanu_twc_13}.
Finally, integration of this p.d.f.~expression yielded the infinite-series expressions for ZF outage probability and ergodic capacity from\cite[Eqs.~(69), (71)]{siriteanu_twc_13}.

\subsection{Limitation of Previous Performance Evaluation Approach}
\label{section_limitation_motivation}

As revealed in\cite{siriteanu_twc_13}\cite{siriteanu_ausctw_14}, the computation of infinite series\cite[Eqs.~(39), (69), (71)]{siriteanu_twc_13} breaks down at much smaller Rician $ K $-factor values than those considered realistic (e.g., averages of lognormal distributions proposed by WINNER II for $ K $ in\cite{winner_d_1_1_2_v_1_2}). 
This is the consequence of the employed expansion of $  {_1\! F_1} (\cdot, \cdot, \sigma) $, whose own computation away from $ \sigma = 0 $ is nontrivial: with increasing $ \sigma $, numerical convergence is increasingly difficult (i.e., slow, resource-intensive) and eventually fails\cite{kim_tvt_12}\cite{muller_nm_01}\cite{koev_mc_06}\cite{grant_tcom_05}.

Note that hypergeometric functions\footnote{Of both scalar and matrix arguments.} and infinite-series expressions have been found to characterize the performance for many other MIMO techniques under many fading types\cite{simon_alouini_book_00}\cite{mckay_tcomm_09}\cite{koev_mc_06}\cite{grant_tcom_05}\cite{ropokis_tcom_10}.
Limitations in computing these infinite-series for MIMO evaluation motivates our search herein for an alternative approach.

\subsection{New Approach and Contribution}

A seldom-considered approach for computing $ {_1\! F_1} (\cdot, \cdot, \sigma) $ follows from the fact that it satisfies, with respect to (w.r.t.) $ \sigma $, a linear differential equation with polynomial coefficients\cite[Eq.~(13.2.1), p.~322]{NIST_book_10}, i.e., 
this function is \textsl{holonomic}\cite[p.~334]{zeilberger_jcam_90}\cite[p.~7]{mallinger_thesis_96}\cite[p.~140]{kauers_book_11}\cite[Section~6.4]{hibi_book_13}. 
Any holonomic function can be computed at some $ \sigma $ by numerically solving its differential equation starting from some $ \sigma_0 $ where the function is either known analytically or can be approximated accurately.

The computation of a holonomic function by numerically solving satisfied differential equations is known as the \textsl{holonomic gradient method} (HGM)\cite{sei_sc_13}\cite{hashiguchi_jma_2013}. 
It has recently been applied in statistics to evaluating the normalizing constant of the Bingham distribution\cite{sei_sc_13} and the cumulative distribution function (c.d.f.) of the dominant eigenvalue of a real-valued Wishart-distributed matrix\cite{hashiguchi_jma_2013}, upon deriving relevant differential equations.
To the best of our knowledge, the HGM has not yet been applied for MIMO performance evaluation, although MIMO performance measures have often been expressed in terms of holonomic special functions and ensuing infinite series\cite{simon_alouini_book_00}\cite{grant_tcom_05}\cite{ropokis_tcom_10}\cite{kim_tvt_12}\cite{siriteanu_twc_13} --- see also\cite{koev_mc_06} and references therein.



The current paper proposes the HGM-based evaluation of the exact MIMO ZF performance under Rician--Rayleigh fading.
Starting from the ZF SNR m.g.f.~expression derived in terms of $ {_1\! F_1} (\cdot, \cdot, \sigma) $ in\cite[Eq.~(31)]{siriteanu_twc_13} and the differential equation satisfied by $ {_1\! F_1} (\cdot, \cdot, \sigma) $\cite[Eq.~(13.2.1), p.~322]{NIST_book_10}, we first deduce differential equations satisfied by the SNR m.g.f..
Their inverse-Laplace transformation yields differential equations satisfied by the SNR p.d.f..

These are shown to enable the accurate HGM-based computation of the SNR p.d.f.~at practical $ K $ values, by starting HGM from an initial p.d.f.~value for $ K \approx 0 $, which can be computed accurately by truncating the available infinite-series p.d.f.~expression from\cite[Eq.~(39)]{siriteanu_twc_13}.
Finally, numerical integration of the HGM output (i.e., the SNR p.d.f.) yields accurately, for the first time, the outage probability and ergodic capacity for MIMO ZF at $ K $ values relevant to WINNER II.

The deduction of differential equations for MIMO performance measures and their HGM-based evaluation may lead to a new framework for MIMO analysis and evaluation under general fading.
As the complexity of MIMO analyses seeking expressions (i.e., explicit representations) of performance measures has been increasing, an ability to derive --- not only manually, as shown herein, but also by computer algebra, as shown in\cite{siriteanu_icc_15} --- differential equations (i.e., implicit representations) for MIMO performance measures and to evaluate them by HGM may be a more general, more straightforward, and more effective alternative.


\subsection{Notation}
\label{section_notation}
The notation defined below follows closely that from\cite{siriteanu_twc_13}.
Thus, scalars, vectors, and matrices are represented with lowercase italics, lowercase boldface, and uppercase boldface, respectively, e.g., $ a $, $ \uh $, and $ \uH $; superscripts $ \cdot^{\cal{T}} $ and $ \cdot^{\cal{H}} $ stand for transpose and Hermitian (i.e., complex-conjugate) transpose; $ [\cdot]_{i,j} $ indicates the $i,j$th element of a matrix; $ \| \uH \|^2 = \sum_{i}^{\NR} \sum_{j}^{\NT} | [\uH]_{i,j} |^2 $ is the squared Frobenius norm of $ \NR \times \NT $ matrix $ \uH $; $ \propto $ stands for `proportional to'; subscripts $ \cdot_{\text{d}} $ and $ \cdot_{\text{r}} $ identify, respectively, the deterministic and random components; subscript $ \cdot_{\text{n}} $ indicates a normalized variable; $ \mathbb{E} \{ \cdot \} $ denotes statistical average; $ (N)_n $ is the Pochhammer symbol, i.e.,  $ (N)_0 = 1 $ and $ (N)_n = N (N + 1) \ldots (N + n - 1) $, $ \forall n > 1 $\cite[p.~xiv]{NIST_book_10}, and $ \,_1F_1(\cdot; \cdot; \cdot) $ is the confluent hypergeometric function\cite[Eq.~(13.2.2), p.~322]{NIST_book_10}.

\subsection{Paper Organization}
Section~\ref{section_system_channel_model} describes the MIMO signal, noise, and channel models.
Section~\ref{section_mgf_pdf} introduces the ZF SNR m.g.f.~and p.d.f.~infinite-series expressions derived in\cite{siriteanu_twc_13}, and discusses difficulties encountered in the computation of the infinite series for the p.d.f.~and ensuing performance measures.
Section~\ref{section_holonomic} defines holonomic functions and deduces from their properties that the SNR m.g.f.~and p.d.f.~are holonomic.
This justifies our search in Section~\ref{section_previous_expressions} for the differential equations they satisfy.
These differential equations are then exploited using the HGM to produce the numerical results shown and discussed in Section~\ref{section_pdf_methods_HGMnumerical}.
Finally, Section~\ref{section_HGM_MIMO_Applications} discusses other possible HGM applications in MIMO evaluation.

\section{Signal, Noise, and Fading Models}
\label{section_system_channel_model}

Herein, the signal, noise, and channel models and assumptions follow closely the ones from\cite{siriteanu_twc_13}.
Thus, we consider uncoded MIMO spatial-multiplexing over a frequency-flat fading channel and assume that there are $ \NT $ and $ \NR $ antenna elements at the transmitter(s) and receiver, respectively, with $ \NT \le \NR $.
Let us denote the number of degrees of freedom as
\begin{eqnarray}
\label{equation_DoFs}
N = \NR - \NT + 1.
\end{eqnarray}
Letting $\ux = [x_1 \, x_2 \, \cdots \, x_{\NT}]^{\cal{T}}$ denote the $ \NT \times 1 $ zero-mean transmit-symbol vector with $ \mathbb{E} \{ \ux \ux^{\cal{H}} \} = \mbI_{\NT} $, the $ \NR \times 1 $ vector with the received signals can be represented as\cite[Eq.~(8)]{paulraj_pieee_04}\cite[Eq.~(1)]{siriteanu_twc_13}:
\begin{eqnarray}
\label{equation_system}
\ur = \sqrt{\frac{E_{\text{s}}}{\NT}} \, \uH \ux + \uv = \sqrt{\frac{E_{\text{s}}}{\NT}} \, \uha x_1 + \sqrt{\frac{E_{\text{s}}}{\NT}} \,  \sum_{k = 2}^{\NT} \uh_k x_k + \uv .
\end{eqnarray}

Above, $ E_{\text{s}}/\NT $ represents the energy transmitted per symbol (i.e., per antenna), so that $ E_{\text{s}} $ is the energy transmitted per channel use.
The additive noise vector $\uv $ is zero-mean, uncorrelated, circularly-symmetric complex-valued Gaussian\cite{paulraj_pieee_04} with variance $ N_0 $ per dimension.
We will also employ its normalized version $ \uv_{\text{n}} = \uv/\sqrt{N_0} $.
We shall employ the per-symbol input SNR, defined as
\begin{eqnarray}
\label{equation_SNR_per_symbol}
 { \Gamma_{\text{s}} }  = \frac{E_{\text{s}}}{N_0} \frac{1}{\NT},
\end{eqnarray}
as well as the per-bit input SNR, which, for a modulation constellation with $ M $ symbols (e.g., $ M $PSK),  is defined as
\begin{eqnarray}
\label{equation_SNR_per_bit}
 { \Gamma_{\text{b}} }  = \frac{\Gamma_{\text{s}} }{\log_2 M} .
\end{eqnarray}

Then, $\uH = (\uha \; \uh_2 \; \ldots \; \uh_{\NT}) $ is the $\NR \times \NT $ complex-Gaussian channel matrix.
Vector $ \uh_k $ comprises the channel factors between transmit-antenna $ k $ and all receive-antennas.
The deterministic (i.e., mean) and random components of $\uH$ are denoted as $\uHd = (\uhda \; \uh_{\text{d},2} \; \ldots \; \uh_{\text{d},\NT}) $ and $\uHr = (\uhra \; \uh_{\text{r},2} \; \ldots \; \uh_{\text{r},\NT})$, respectively, so that $ \uH = \uHd + \uHr $.
If $ [\uHd]_{i,j} = 0 $ then $ |\left[ \uH \right]_{i,j}| $ has a Rayleigh distribution; otherwise, $ |\left[ \uH \right]_{i,j}| $ has a Rician distribution\cite{simon_alouini_book_00}.
Typically, the channel matrix for Rician fading is written as\cite{paulraj_pieee_04}
\begin{eqnarray}
\label{equation_channelH}
\uH =  \uHd + \uHr = \sqrt{\frac{K}{K+1}} \, \uHdn + \sqrt{\frac{1}{K+1}} \, \uHrn,
\end{eqnarray}
where, for normalization purposes\cite{loyka_twc_09}, it is assumed that 
\begin{eqnarray}
\label{equation_my_normalization}
\| \uHdn \|^2 = \mathbb{E} \{ \| \left[ \uHrn \right] \|^2 \} = \NT \NR,
\end{eqnarray}
so that $ \mathbb{E} \{ \| \uH \|^2 \} = \NT \NR $.
Power ratio
\begin{eqnarray}
\label{equation K_definition}
K = \frac{\| \uHd \|^2}{\mathbb{E} \{ \| \uHr \|^2 \} } = \frac{ \frac{K}{K+1} \| \uHdn \|^2}{\frac{1}{K+1} \mathbb{E} \{ \| \uHrn \|^2 \} }
\end{eqnarray}
is the Rician $ K $-factor: $ K = 0 $ yields Rayleigh fading for all elements of $ \uH $; $ K \neq 0 $ yields Rician fading if $ \uHdn \neq \mzero $.

In\cite{siriteanu_twc_13}, we partitioned into the column with the fading gains that affect the intended stream, e.g., Stream 1, and the matrix  columns with the fading gains that affect the interfering streams, i.e.,
\begin{eqnarray}
\label{equation_partitioned_H}
\uH = ( \uha \; \; \uHb ) = ( \uhad \; \; \uHdb ) + ( \uhar \; \; \uHrb ),
\end{eqnarray}
and assumed, for analysis tractability, that $ \uhad \neq \mzero $ and $ \uHdb = \mzero $, i.e., Rician--Rayleigh fading.
Then, we can write
\begin{eqnarray}
\label{equation_mu_norm}
\| \uhad \|^2 & = & \| ( \uhad \; \; \mzero_{\NR \times (\NT - 1)} ) \|^2 = \| \uHd \|^2 \nonumber \\ & = & \frac{K}{K + 1} \NR \NT.
\end{eqnarray}

As in\cite{gore_cl_02}\cite{kiessling_spawc_03}, for tractability, we assume zero receive-correlation and we allow for nonzero transmit-correlation whereby all conjugate-transposed rows of $ \uHrn $ have distribution $ {\cal{CN}} (\mzero, \uRT ) $. 
Consequently, all conjugate-transposed  rows of $ \uHr $ have distribution $ {\cal{CN}} (\mzero, \uRTK = \frac{1}{K + 1 } \uRT  ) $.

It can be shown that normalization $ \mathbb{E} \{ \| \left[ \uHrn \right] \|^2 \} = \NT \NR $ is equivalent with
\begin{eqnarray}
\label{equation_RT_cond}
 \sum_{i = 1}^{\NT} [\uRT]_{i,i} = \NT.
\end{eqnarray}
Because the diagonal elements of $ \uRT $ need not be equal, our analysis applies also for distributed transmitters. 
Nevertheless, for simplicity, numerical results are shown herein only for the case $ [\uRT]_{i,i} = 1 $, $ \forall i $, i.e., for collocated transmitters.

The elements of $ \uRT $ can be computed from the azimuth spread (AS) as shown in\cite[Section~VI.A]{siriteanu_tvt_11} for WINNER II, i.e., Laplacian, power azimuth spectrum. Note that WINNER II has modeled both $ K $ (in dB) and AS (in degrees) as random variables with scenario-dependent lognormal distributions\cite{winner_d_1_1_2_v_1_2}.
Herein, we show results for $ K $ and AS set to their averages for WINNER II indoor scenario A1.


\section{Infinite-Series Expressions for MIMO ZF SNR M.G.F.~and P.D.F.}
\label{section_mgf_pdf}

\subsection{MIMO ZF and Its SNR M.G.F.~for Rician--Rayleigh Fading}
\label{section_SNR_intro}

For the received-signal vector from~(\ref{equation_system}), ZF means identifying the closest modulation constellation symbol for each element of the vector\cite[Eq.~(22)]{paulraj_pieee_04}
\begin{eqnarray}
\label{equation ZF_for_perfect_CSI}
\sqrt{\frac{ \NT }{ E_{\text{s}} }} \left[ \uH^{\cal{H}} \uH \right]^{-1} \uH^{\cal{H}} \, \ur
= \ux +\frac{ 1 }{  \sqrt{ { \Gamma_{\text{s}} }  }} \left[ \uH^{\cal{H}} \uH \right]^{-1} \uH^{\cal{H}} \uv_{\text{n}}.
\end{eqnarray}
Then, the ZF SNR for Stream 1 is given by\cite{winters_tcom_94}
\begin{eqnarray}
\label{equation_gammak_perfect_CSI}
\gamma_1 = \frac{  { \Gamma_{\text{s}} }  }{\left[(\uHH \uH )^{-1}\right]_{1,1}}.
\end{eqnarray}
Its m.g.f.~is defined as\cite[Eq.~(1.2)]{simon_alouini_book_00}
\begin{eqnarray}
\label{equation_mgf_definition_1}
M_{\gamma_1}(s) = \mathbb{E} \{ e^{s \gamma_1} \} = \int_0^{\infty} e^{s t} p_{\gamma_1} (t)  \mathrm{d} t,
\end{eqnarray}
where $ \mathbb{E} \{ \cdot \} $ stands for mean, and $ p_{\gamma_1} (t, a) $ is the p.d.f.~of $ \gamma_1 $.
Thus, the m.g.f.~is related to the Laplace transform\cite[Eq.~(1.14.17)]{NIST_book_10} $ L_{\gamma_1}(s, a) $ of the p.d.f.~by the sign change
\begin{eqnarray}
\label{equation_mgf_Laplace_rel}
M_{\gamma_1}(-s) = L_{\gamma_1}(s) = \int_0^{\infty} e^{-s t} p_{\gamma_1} (t)  \mathrm{d} t.
\end{eqnarray}

If we define as in\cite[Eqs.~(18), (23)]{siriteanu_twc_13}, respectively,
\begin{eqnarray}
\label{equation_Gamma}
\Gamma_1 & = & \frac{ { \Gamma_{\text{s}} } }{\RTinvaa} \propto \frac{ { \Gamma_{\text{s}} } }{K+1}, \\
\label{equation_a_first}
a & = & \RTinvaa \| \uhad \|^2 \propto K \NR \NT,
\end{eqnarray}
then, for MIMO ZF under Rician--Rayleigh fading, we can write the following exact expression for the m.g.f.~of the SNR of the Rician-fading Stream 1, from\cite[Eq.~(31)]{siriteanu_twc_13}:
\begin{eqnarray}
\label{equation_gamma1_mgf_final}
M_{\gamma_1}(s, a) = \frac{1}{\left( 1 - \Gamma_1 s \right)^N} {_1\! F_1} \left(N; \NR; a \frac{\Gamma_1 s}{1 - \Gamma_1 s} \right),
\end{eqnarray}
where $ {_1\! F_1}(N; \NR; \sigma)  $ is the confluent hypergeometric function of scalar argument $ \sigma $\cite[Ch.~13]{NIST_book_10}.

For the SNR m.g.f.~in~(\ref{equation_gamma1_mgf_final}) we have added $ a \propto K $ as variable, and, hereafter, also for the SNR p.d.f., because for  p.d.f.~evaluation  we shall apply HGM, which requires differential equations w.r.t.~$ a $ --- they are deduced from~(\ref{equation_gamma1_mgf_final}) further below, in Section~\ref{section_inverse_laplace_2}. 
Hereafter in this Section, we show the infinite series obtained for the SNR p.d.f.~in\cite{siriteanu_twc_13}, and explain its numerical convergence difficulties with increasing $ a $ (i.e., $ K $).


\subsection{Infinite Series for SNR M.G.F.~and P.D.F.~for Rician--Rayleigh Fading}
\label{section_infinite_sums_twc_13}

The confluent hypergeometric function is well-known to have the infinite-series expansion\cite[Eq.~(13.2.2), p.~322]{NIST_book_10}
\begin{eqnarray}
\label{equation_1F1_series}
{_1\! F_1}(N; \NR; \sigma) = \sum_{n = 0}^{\infty} { \frac{\left( N \right)_n}{\left( \NR \right)_n} \frac{\sigma^n}{n!} } {= \sum_{n = 0}^{\infty}  A_n(\sigma)}
\end{eqnarray}
around $ \sigma = 0 $, as can be readily proved from the integral expression for $ {_1\! F_1}(N; \NR; \sigma) $ in\cite[Eq.~(13.4.1), p.~326]{NIST_book_10}.

Using~(\ref{equation_1F1_series}), we rewrote~(\ref{equation_gamma1_mgf_final}) in\cite[Eq.~(37)]{siriteanu_twc_13} as the infinite series
\begin{eqnarray}
\label{equation_gamma1_mgf_final_again}
M_{\gamma_1}(s, a) =  \sum_{n = 0}^{\infty} A_n(a) \sum_{m = 0}^{n} {n \choose m}   {\frac{ (-1)^{m} }{(1 - s \Gamma_1)^{N + n - m}}}.
\end{eqnarray}
Its inverse-Laplace transformation has yielded for the SNR p.d.f.~the infinite series\footnote{Note that, for $ K \neq 0 $, i.e., for Rician--Rayleigh fading,~(\ref{equation_gamma1_mgf_final_again}) and~(\ref{equation_gamma1_pdf_final}) reveal that the distribution of the ZF SNR is an infinite linear combination of Gamma distributions.}\cite[Eq.~(39)]{siriteanu_twc_13}
\begin{eqnarray}
\label{equation_gamma1_pdf_final}
p_{\gamma_1} (t, a) & = & \sum_{n = 0}^{\infty} A_n(a) \sum_{m = 0}^{n} {n \choose m}  \nonumber \\ 
&& \times 
\frac{ (-1)^{m} t^{N + n - m - 1}  e^{-t/\Gamma_1}}{[(N + n - m) - 1]! \,  \Gamma_1^{N + n - m}}.
\end{eqnarray}

For the outage probability and ergodic capacity (in bits per channel use --- bpcu), which are defined as\footnote{In~(\ref{equation_Po_definition}), $ \gamma_{1,\text{th}} $ is the threshold SNR.}
\begin{eqnarray}
\label{equation_Po_definition}
P_{\text{o}} (\gamma_{1,\text{th}}, a) = \text{Probability} ( \gamma_1 \le  \gamma_{1,\text{th}}) = \int_{0}^{\gamma_{1,\text{th}}} p_{\gamma_1 } (t, a) \mathrm{d} t, \\
\label{equation_capacity_ergodic_inf_sum}
C(a) = \mathbb{E}_{\gamma_1 } \{ C(\gamma_1, a) \} = \int_{0}^{\infty} \log_2 (1 + t) p_{\gamma_1} (t, a) \mathrm{d} t. \quad \; \;
\end{eqnarray}
analytical integration of the infinite-series p.d.f.~expression~(\ref{equation_gamma1_pdf_final}) has yielded the infinite-series expressions in\cite[Eqs.~(69), (71)]{siriteanu_twc_13}, respectively.

\subsection{Expressions for Rayleigh-Only Fading}
\label{section_Rayleigh_fading}

For the special case of Rayleigh-only fading, because $ a = 0 $, only the term for $ n = m = 0 $ remains in~(\ref{equation_gamma1_mgf_final_again}) and~(\ref{equation_gamma1_pdf_final}), i.e.,
\begin{eqnarray}
\label{equation_gamma1_mgf_Ray}
M_{\gamma_1}(s, 0) & = & {\frac{ 1 }{(1 - s \Gamma_1)^{N }}}, \\
\label{equation_gamma1_pdf_Ray}
p_{\gamma_1} (t, 0) & = & \frac{t^{N - 1}  e^{-t/\Gamma_1}}{(N - 1)! \,  \Gamma_1^{N }}, \quad t \ge 0,
\end{eqnarray}
so that the ZF SNR is Gamma-distributed\cite{gore_cl_02}\cite{kiessling_spawc_03}.
Then,~(\ref{equation_Po_definition}) and~(\ref{equation_capacity_ergodic_inf_sum}) yield\footnote{The finite- and infinite-limit integrals involved in these expressions can be computed accurately numerically.}
\begin{eqnarray}
\label{equation_Po_Rayleigh}
P_{\text{o}} (\gamma_{1,\text{th}}, 0) = \frac{1}{(N - 1)!\Gamma_1^{N}}
\int_{0}^{\gamma_{1,\text{th}}} t^{N - 1} e^{-t/\Gamma_1} \mathrm{d} t, \quad \quad \; \; \\
\label{equation_C_Rayleigh}
C(0) = \frac{1}{\ln 2}  \frac{ 1 }{ (N - 1)! } \frac{ 1 }{ \Gamma_1^{N} }  \int_{0}^{\infty} \ln (1 + t) t^{N - 1} e^{-t/\Gamma_1} \mathrm{d} t.
\end{eqnarray}

\subsection{Difficulties in Computing the Derived Infinite Series}
\label{section_SNR_MGF_PDF_difficult}

As mentioned in the Introduction, we proved analytically in\cite{siriteanu_ausctw_14} that the infinite series~(\ref{equation_gamma1_pdf_final}), along with the ensuing infinite series for the outage probability and ergodic capacity, converge everywhere. 
However, they cannot be computed (by truncation) accurately, or even at all, with increasing $ K $ --- see\cite[Sections~V.F, VI.C]{siriteanu_twc_13}\cite{siriteanu_ausctw_14} for discussion and results.

\begin{figure}[t]
\begin{center}
\includegraphics[width=3.5in]
{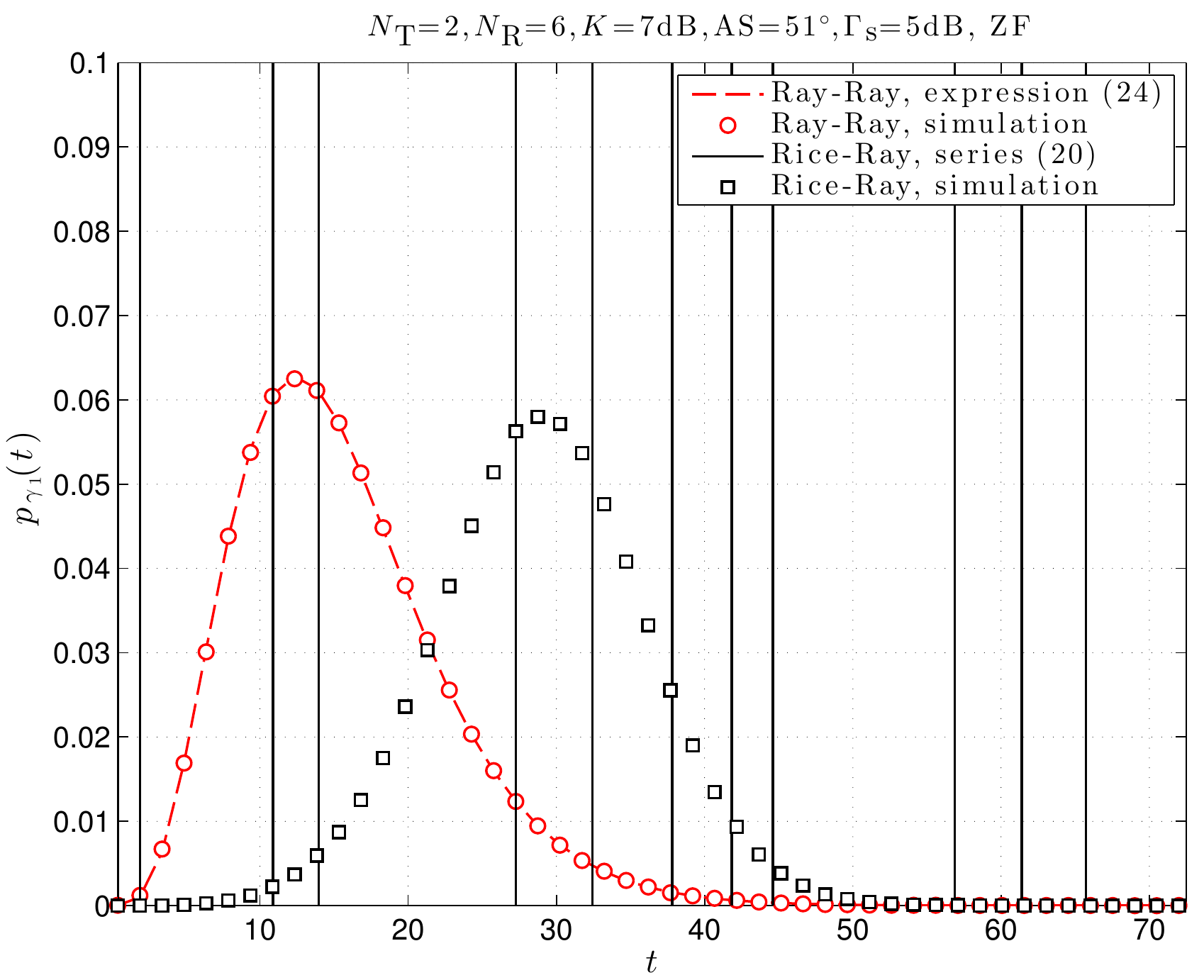}
\caption{P.d.f.~of the SNR (in linear units) for Stream 1, for $ \NR = 6 $, $ \NT = 2 $, $ K = 7 $ dB, $ \text{AS}= 51^\circ $, $  { \Gamma_{\text{s}} }  = 5 $~dB.
For Rayleigh fading: from Monte Carlo simulation and expression~(\ref{equation_gamma1_pdf_Ray}).
For Rician fading: from simulation and attempt to employ series~(\ref{equation_gamma1_pdf_final}); the latter produced the vertical black lines (which connect the points of extreme positive and negative values resulting from series truncation).}
\label{figure_ZF_SNR_pdf_Rayleigh_Rice_Fixed_AS_K_70_A1_NT2_NR6}
\end{center}
\end{figure}

This limitation is also illustrated herein for the SNR p.d.f.~infinite series in~(\ref{equation_gamma1_pdf_final}) in Fig.~\ref{figure_ZF_SNR_pdf_Rayleigh_Rice_Fixed_AS_K_70_A1_NT2_NR6}, for $ \NR = 6 $, $ \NT = 2 $. We have set $ K = 7 $~dB and $ \text{AS} = 51^\circ $, i.e., the average $ K $ and AS for WINNER II scenario A1 (indoor office/residential)\cite[Table~I]{siriteanu_tvt_11}\cite{winner_d_1_1_2_v_1_2}.
On one hand, results for Rayleigh-only fading (identified in legend with \texttt{Ray}--\texttt{Ray}) reveal agreement between  expression~(\ref{equation_gamma1_pdf_Ray}) and  Monte Carlo simulation.
On the other hand, for Rician--Rayleigh fading (identified in legend with \texttt{Rice}--\texttt{Ray}), the results from series~(\ref{equation_gamma1_pdf_final}) are not usable.
For $ \NR = 6 $ and $ \NT = 2 $, we have been able to accurately compute $ p_{\gamma_1} (t) $, and, thus, the outage probability and ergodic capacity, only up to $ K \approx 1.5 $~dB, as depicted in\cite[Fig.~2]{siriteanu_ausctw_14}.
This is because, by increasing $ K $, i.e., $ a $, we move $ \sigma $ further from the origin of expansion~(\ref{equation_1F1_series}).
Finally, we have also found that increasing $ \NT $ to $ 6 $ decreases the value of $ K $ that still yields accurate results to $ -3  $~dB\cite[Figs.~1, 2]{siriteanu_ausctw_14}, which is explained by~(\ref{equation_a_first}).



Because infinite-series truncation cannot help compute MIMO ZF performance measures for relevant fading parameter values, we pursue next a novel HGM-based approach.

\section{Holonomic Functions and the Holonomic Gradient Method (HGM)}
\label{section_holonomic}

\subsection{Differential Equation for $ {_1\! F_1}(N; \NR; \sigma) $}
\label{section_1F1_diff_eq}

It is known that $ {_1\! F_1} (N; \NR; \sigma) $ satisfies the second-order ordinary differential equation with polynomial coefficients\cite[Eq.~(13.2.1), p.~322]{NIST_book_10}
\begin{eqnarray}
\label{equation_1F1_diff_eq}
&& \sigma \cdot {_1\!F_1^{(2)}} (N; \NR; \sigma) + (\NR - \sigma) \cdot {_1\!F_1^{(1)}} (N; \NR; \sigma) \nonumber \\
&& \quad - N \cdot {_1\! F_1} (N; \NR; \sigma) = 0,
\end{eqnarray}
where $ {_1\!F_1^{(k)}} (N; \NR; \sigma) $ stands for the $ k $th derivative w.r.t.~the sole variable $ \sigma $.

In general, a function is called \textsl{holonomic} if it satisfies, w.r.t.~each variable, an ordinary differential equation with polynomial coefficients\cite[Section~2]{sei_sc_13}. 
Thus, the confluent hypergeometric function $ {_1\! F_1} (N; \NR; \sigma) $ is holonomic because it satisfies~(\ref{equation_1F1_diff_eq}).
Simpler examples of holonomic functions are the polynomial and exponential-polynomial functions\cite[Section~2.5]{zeilberger_jcam_90}.

Let us now introduce the HGM-based computation of a holonomic function on the example of $ {_1\! F_1} (N; \NR; \sigma) $.
First,~(\ref{equation_1F1_diff_eq}) can be recast as the system of differential equations
\begin{eqnarray}
\label{equation_diff_eq_system_1F1_1}
   \partial_{\sigma} 
\begin{pmatrix}
{_1\! F_1} (N; \NR; \sigma) \\
{_1\! F_1^{(1)}} (N; \NR; \sigma)
 \end{pmatrix}
 =
 \begin{pmatrix}
  0 & 1 \\
  \frac{N}{\sigma} & 1 - \frac{\NR}{\sigma}
 \end{pmatrix}
\begin{pmatrix}
{_1\! F_1} (N; \NR; \sigma) \\
{_1\! F_1^{(1)}} (N; \NR; \sigma)
 \end{pmatrix}. \nonumber
 \end{eqnarray}
If we denote as $ \uf(\sigma) $ the vector made of functions $ {_1\! F_1} (N; \NR; \sigma) $ and ${_1\! F_1^{(1)}} (N; \NR; \sigma)  $, and as $ \uF(\sigma) $ the $ 2 \times 2 $ matrix on the right-hand side above, which is also known as the \textsl{companion} matrix, then, we have, more compactly,
\begin{eqnarray}
\label{equation_diff_eq_system_1F1}
   \partial_{\sigma} 
\uf(\sigma)
 = \uF(\sigma) \uf(\sigma),
\end{eqnarray}
whose left-hand side is the \textsl{gradient} of $ \uf(\sigma) $ w.r.t.~$ \sigma $.

\subsection{HGM-Based Computation of Holonomic Function}

Let us assume that initial conditions $ {_1\! F_1}(N; \NR; \sigma_0) $ and $ {_1\!F_1^{(1)}}(N; \NR; \sigma_0) $ are known  for some $ \sigma_0 $. 
Then, $ {_1\! F_1}(N; \NR; \sigma) $ can be computed for any $ \sigma $ by numerically solving\footnote{E.g., with the \texttt{ode} function, in \texttt{MATLAB}.}~(\ref{equation_diff_eq_system_1F1}) between $ \sigma_0 $ and $ \sigma $.
Because $ \sigma $ appears in $ \uF(\sigma) $ denominators, one cannot use $ \sigma_0 = 0 $, for which it is known analytically, from~(\ref{equation_1F1_series}) and\cite[Eq.~(13.3.15), p.~325]{NIST_book_10}, that $ {_1\! F_1} (N; \NR; 0) = 1 $ and $ {_1\! F_1^{(1)}} (N; \NR; 0) = \frac{N}{\NR} $.
Thus, initial conditions $ {_1\! F_1} (N; \NR; \sigma_0) $ and $ {_1\! F_1^{(1)}} (N; \NR; \sigma_0) = \frac{N}{\NR} \, {_1\! F_1} (N + 1; \NR + 1; \sigma_0) $\cite[Eq.~(13.3.15), p.~325]{NIST_book_10} have to be obtained numerically by truncation of series~(\ref{equation_1F1_series}) for some $ \sigma_0 > 0 $.
Nevertheless, since $ \sigma_0 $ can be selected arbitrarily small, highly-accurate computation of the initial condition $ \uf(\sigma_0) $ is possible based on the infinite-series expression of $ {_1\! F_1}(N; \NR; \sigma) $ from~(\ref{equation_1F1_series}).


The entire procedure is summarized below\cite[Section~2.1]{sei_sc_13}:
\begin{itemize}
\item First, compute accurate initial conditions, i.e., $ {_1\! F_1} (N; \NR; \sigma_0) $ and $ {_1\! F_1^{(1)}} (N; \NR; \sigma_0) $, for a sufficiently-small $ \sigma_0 > 0 $, by truncating infinite series~(\ref{equation_1F1_series}).
\item Then, solve numerically the system of differential equations~(\ref{equation_diff_eq_system_1F1}) between $ \sigma_0 $ and $ \sigma $.
\end{itemize}
This computation procedure (applicable to any holonomic function) is referred to as the \textsl{holonomic gradient method}\cite{sei_sc_13}\cite{hashiguchi_jma_2013} (HGM) because, after starting from an initial condition, the \textsl{holonomic} function is computed by updating its \textsl{gradient}.


%
%
%

\subsection{ZF SNR M.G.F.~and P.D.F.~Are Holonomic Functions}

Holonomic functions satisfy the following properties:
\begin{enumerate}
\item If $ f(x) $ is a polynomial then $ 1/f(x) $ is holonomic\cite[Prop.~2.1]{zeilberger_jcam_90}.
\item If $ f(x) $ is holonomic and $ h(x) $ is rational then $ f(h(x)) $ is holonomic\cite[Th.~1.4.2, p.~16]{mallinger_thesis_96}.
\item If $ f(x) $ and $ g(x) $ are holonomic then $ f(x) \, g(x) $ is holonomic\cite[Prop.~3.2]{zeilberger_jcam_90}.
\item If $ f(x) $ is holonomic then its Fourier transform is holonomic\cite[p.~337]{zeilberger_jcam_90}.
\end{enumerate}
Properties 1 and 2 reveal that in the SNR m.g.f.~expression~(\ref{equation_gamma1_mgf_final}),
the term $ {1}/{\left( 1 - \Gamma_1 s \right)^N} $ is holonomic w.r.t.~$ s $, and the term $ {_1\! F_1} \left(N; \NR; a \frac{\Gamma_1 s}{1 - \Gamma_1 s} \right) $ is holonomic w.r.t.~$ s $ and $ a $.
Then, Properties 3 and 4 prove the following lemma.

\begin{lemma}
\label{property_holonomic_mgf_pdf}
The SNR m.g.f.~$ M_{\gamma_1}(s, a) $ is holonomic w.r.t.~both $ s $ and~$ a $, i.e., it must satisfy ordinary differential equations with polynomial coefficients w.r.t.~both $ s $ and~$ a $.
Then, the SNR p.d.f.~$ p_{\gamma_1} (t, a) $ is holonomic w.r.t.~both $ t $ and~$ a $, i.e., it must satisfy ordinary differential equations with polynomial coefficients w.r.t.~both $ t $ and~$ a $.
\end{lemma}

\subsection{ZF Performance Analysis and Evaluation Procedure}
\label{remark_explanation_steps}

The rest of this work is devoted to deriving relevant differential equations and to using them for HGM-based ZF performance evaluation. The procedure is as follows:
\begin{enumerate}
\item Deduce differential equations --- known to exist from Lemma~\ref{property_holonomic_mgf_pdf} --- satisfied by $ M_{\gamma_1}(s, a) $ w.r.t.~both $ a $ and~$ s $, as well as by $ p_{\gamma_1} (t, a) $ w.r.t.~both $ a $ and~$ t $. Considering both variables in each pair is necessary because, as we shall see, the partial derivatives w.r.t.~the two variables appear together in differential equations for $ M_{\gamma_1}(s, a) $ and for $ p_{\gamma_1} (t, a) $.
\item Exploit the deduced differential equations for $ p_{\gamma_1} (t, a) $ to compute the p.d.f.~by HGM at relevant values of $ K $.
\item Integrate numerically the p.d.f.~produced by the HGM, based on~(\ref{equation_Po_definition}), (\ref{equation_capacity_ergodic_inf_sum}), to compute  the outage probability and ergodic capacity.
\end{enumerate}

\section{Differential Equations for SNR M.G.F., P.D.F.}
\label{section_previous_expressions}

\subsection{M.G.F.~and P.D.F.~Variable Scaling}
\label{section_SNR_MGF_PDF_difficult_1}


In order to simplify notation and derivations hereafter, let us denote the m.g.f.~$ M_{\gamma_1}(s, a) $ and the p.d.f.~$ p_{\gamma_1} (t, a) $ for $ \Gamma_1 = 1 $ as $ M(s, a) $ and $ p (t, a) $, respectively. 
Now, by definition, we have
\begin{eqnarray}
\label{equation_mgf_normalized}
M(s, a) = \int_0^{\infty} e^{s t} p (t, a) \mathrm{d} t.
\end{eqnarray}
Then, because
\begin{eqnarray}
\label{equation_mgf_unnormalized}
M_{\gamma_1}(s, a) & = & M(s \Gamma_1, a) = \int_0^{\infty} e^{s \Gamma_1 t} p (t, a)  \mathrm{d} t \nonumber \\ & = & \int_0^{\infty} e^{s y} \frac{1}{\Gamma_1} p \bigg(\frac{y}{\Gamma_1}, a \bigg)  \mathrm{d} y, \nonumber
\end{eqnarray} 
the p.d.f.~$ p_{\gamma_1} (t, a) $ of $ \gamma_1 $ for any $ \Gamma_1 $ can be obtained from $ p (t, a) $ as follows:
\begin{eqnarray}
\label{equation_pdf_relation}
p_{\gamma_1} (t, a) = \frac{1}{\Gamma_1} p \bigg(\frac{t}{\Gamma_1}, a \bigg) .
\end{eqnarray}

Hereafter, we first derive differential equations for $ M(s, a) $ w.r.t.~both $ s $ and $ a $.
From them we then deduce differential equations for $ p(t, a) $ w.r.t.~both $ t $ and $ a $.
They will help compute, by HGM, the function $ p (t, a) $ at desired values of $ t $ and $ a $ (i.e., $ K $).
Finally, the transformation from~(\ref{equation_pdf_relation}) will return the value of the SNR p.d.f.~$ p_{\gamma_1} (t, a) $, for any $ \Gamma_1 $.

We proceed below as follows. First, we deduce the differential equation w.r.t.~$ s $ for $ M(s, a) $. From it we then derive the differential equation w.r.t.~$ t $ for $ p(t, a) $, which, in turn, helps derive the differential equation w.r.t.~$ a $ for $ p(t, a) $.

\subsection{Differential Equation w.r.t.~$ s $ for $ M(s, a) $}
\label{section_find_ode_invert_order}

Based on~(\ref{equation_gamma1_mgf_final}) and~(\ref{equation_mgf_unnormalized}) we can write
\begin{eqnarray}
\label{equation_gamma1_mgf_final_scaled_x}
M(s, a) = \frac{1}{\left( 1 - s \right)^N} {_1\! F_1} \left(N; \NR;  \frac{a s}{1 - s} \right).
\end{eqnarray}
In Appendix~\ref{section_derivation_SNR_MGF_ODEs}, manipulation and differentiation w.r.t.~$ s $ of~(\ref{equation_gamma1_mgf_final_scaled_x}) followed by substitution into the differential equation for $ {_1\! F_1} \left(N; \NR;  \sigma \right) $ from~(\ref{equation_1F1_diff_eq}) have yielded the following differential equation w.r.t.~$ s $ for $ M(s, a) $, in~(\ref{equation_MGF_diff_eq_wrt_s_appendix}):
\begin{eqnarray}
\label{equation_MGF_diff_eq_wrt_s_final_main}
\bigg( s (1 - s)^2 \partial_s^2 - [ 2 (N + 1) s (1 - s) \!- \!(1 - s) \NR + a s ] \partial_s \nonumber \\ + N [ (N + 1) s -\NR - a ]  \bigg) M(s, a) = 0.
\end{eqnarray} 
Because $ s^l $ appears in front of $ \partial_s^k $ in~(\ref{equation_MGF_diff_eq_wrt_s_final_main}), the corresponding differential equation for $ p(t, a) $ cannot be obtained by inverse-Laplace transform. 
Therefore, we shall first employ the following order-changing rule, which can readily be deduced from\cite[Th.~6.1.2 (Leibniz Formula), p.~282]{hibi_book_13}\cite[Th.~1.1.1, p.~3]{saito_book_11}. 



\begin{proposition}
\begin{eqnarray}
\label{equation_liebniz}
s^l \partial_s^k = \! \! \sum_{m=0}^{\min(l, k)} \frac{(-1)^m{(l-m+1)_{m}  (k-m+1)_{m}}}{m!}  \partial_s^{k-m} s^{l-m}. \nonumber
\end{eqnarray}
\end{proposition}





The above general rule yields the following particular rules
\begin{eqnarray}
\label{equation_partials_s_rules_1}
s \partial_s & = & \partial_s s - 1, \\
\label{equation_partials_s_rules_2}
s \partial_s^2 & = & \partial_s^2 s - 2 \partial_s, \\
\label{equation_partials_s_rules_3}
s^2 \partial_s & = & \partial_s s^2 - 2 s, \\
\label{equation_partials_s_rules_4}
s^2 \partial_s^2 & = & \partial_s^2 s^2 - 4 \partial_s s + 2, \\
\label{equation_partials_s_rules_5}
s^3 \partial_s^2 & = & \partial_s^2 s^3 - 6 \partial_s s^2 + 6 s,
\end{eqnarray}
which, when applied in~(\ref{equation_MGF_diff_eq_wrt_s_final_main}), yield for $ M(s, a) $ the following differential equation w.r.t.~$ s $:
\begin{eqnarray}
\label{equation_diff_eq_m_wrt_s_processed}
&&[\partial_s^2 s^3 - 2 \partial_s^2 s^2 + \partial_s^2 s +(2N - 4) \partial_s s^2 \nonumber \\ 
&& \quad + \left( 6 - 2 N - \NR - a \right) \partial_s s  + (\NR - 2) \partial_s \nonumber \\ 
&& \quad + (N - 1) (N - 2) s \nonumber \\ && \quad  + (N - 1) (2 - \NR - a) ] M(s, a) = 0.
\end{eqnarray}
Unlike Eq.~(\ref{equation_MGF_diff_eq_wrt_s_final_main}), Eq.~(\ref{equation_diff_eq_m_wrt_s_processed}) can be employed based on the inverse-Laplace transform to deduce the differential equation w.r.t.~$ t $ for $ p(t, a) $, as shown next.

\subsection{Differential Equation w.r.t.~$ t $ for $ p(t, a) $}
\label{section_pdf_eq_wrt_t}

The following proposition helps transform an expression whereby the operator $ \partial_s^k $ is applied to the product $ s^l M(s, a) $ into a differential equation for $ p(t, a)  $ w.r.t.~$t$.
Hereafter, $ p^{(l)} (t, a) $ stands for the $ l $th partial derivative w.r.t.~$ t $ of $ p (t, a) $.

\begin{proposition}
\label{lemma_derivatives}
The integral $ \int_0^{\infty} e^{s t} \left[ t^k p^{(l)} (t, a)  \right] \mathrm{d} t $, which represents the Laplace transform of $ t^k p^{(l)} (t, a) $ for argument $ - s $, is given by:
\begin{eqnarray}
\begin{cases}
\label{equation_diff_integral_rule_1}
(-1)^l \partial_s^k [ s^l M(s, a)] \\ \quad + \sum_{m = k + 1}^{l} (-1)^{m} \, p^{(l - m)}(0+, a) \\ \quad \quad \times \frac{(m-1)!}{(m - k - 1)!} s^{m- k - 1}, & \! \! \! l \ge 1, \\
\partial_s^k M(s, a) , & \! \! \! l = 0.
\end{cases}
\end{eqnarray}
\end{proposition}


\IEEEproof{Follows from the well-known Laplace-transform property for higher-order derivatives from\cite[Eq.~(1.14.29), p.~28]{NIST_book_10} and the sign change from~(\ref{equation_mgf_Laplace_rel}).}


Applying~(\ref{equation_diff_integral_rule_1}) for the terms in~(\ref{equation_diff_eq_m_wrt_s_processed}) yields the following Laplace-transform pairs:
\begin{eqnarray}
\label{equation_diff_eq_m_wrt_s_partial}
&& \partial_s^2 s^3  M(s, a) + 2! \, p(0+, a)  \leftrightarrow \nonumber \\
&& \quad \quad \quad \quad -  t^2 p^{(3)} (t, a) \nonumber \\
&& - 2 \partial_s^2 s^2 M(s, a) \leftrightarrow \nonumber \\
&& \quad \quad \quad \quad  - 2   t^2 p^{(2)} (t, a) \nonumber \\
&& \partial_s^2 s M(s, a) \leftrightarrow \nonumber \\
&& \quad \quad \quad \quad -  t^2 p^{(1)} (t, a) \nonumber \\
&& (2N - 4) \partial_s s^2 M(s, a) + (2N - 4) \, p(0+, a) \leftrightarrow \nonumber \\
&& \quad \quad \quad \quad \left(2  N - 4\right)   t \, p^{(2)} (t, a) \nonumber \\
&& \left( 6 - 2 N - \NR - a \right) \partial_s s M(s, a) \leftrightarrow \nonumber \\
&& \quad \quad \quad \quad  -\left( 6 - 2 N - \NR - a \right)   t \, p^{(1)} (t, a)  \nonumber \\
&& (\NR - 2) \partial_s M(s, a) \leftrightarrow \nonumber \\
&& \quad \quad \quad \quad  \left( \NR - 2\right)  t \, p (t, a) \nonumber \\
&& (N - 1) (N - 2) s M(s, a) + (N - 1) (N - 2) p(0+, a) \leftrightarrow\nonumber \\
&& \quad \quad \quad \quad   - \left(N - 1\right)  \left(N - 2\right)  p^{(1)} (t, a)  \nonumber \\
&& (N - 1) (2 - \NR - a) M(s, a) \leftrightarrow \nonumber \\
&& \quad \quad \quad \quad (N - 1) (2 - \NR - a) \, p(t, a). \nonumber
\end{eqnarray}
Summing the left-hand-side terms (i.e., the $ s $-domain terms) of the above transform pairs and accounting for~(\ref{equation_diff_eq_m_wrt_s_processed}) yield the constant $ N (N - 1) p(0+, a) $. 
However, this constant is zero because~(\ref{equation_pdf_relation}) above and~(\ref{equation_p__gamma1_0_plus}) in Appendix~\ref{section_initial_condition} yield
\begin{eqnarray}
\label{equation_p_0_plus}
p(0+, a) = 
\begin{cases}
 {_1\! F_1}(N; \NR; -a), & N = 1, \\
0, & N > 1.
\end{cases}
\end{eqnarray}

Then, by the uniqueness of the Laplace transform, the right-hand-side terms (i.e., the $ t $-domain terms) of the above transform pairs also sum to $ 0 $, i.e.,
\begin{eqnarray}
\label{equation_ode_pdf_wrt_t_first}
&& -  t^2 p^{(3)} (t, a) - 2   t^2 p^{(2)} (t, a) -  t^2 p^{(1)} (t, a) \nonumber \\
&& \quad + (2  N - 4)   t p^{(2)} (t, a) -( 6 - 2 N - \NR - a )  t \, p^{(1)} (t, a) \nonumber \\ 
&& \quad + ( \NR - 2)  t \, p (t, a) - (N - 1)  (N - 2)  p^{(1)} (t, a) \nonumber \\
&& \quad + (N - 1) (2 - \NR - a) \, p(t, a) = 0,
\end{eqnarray}
which can be rewritten as the differential equation w.r.t.~$ t $
\begin{eqnarray}
\label{equation_f3}
&& p^{(3)} (t, a) = {\frac{ ( \NR - 2)  t + (N - 1) (2 - \NR - a) }{t^2}} p(t, a) \nonumber \\
 & & \; {-\frac{ t^2 + \! ( 6 -\!  2 N - \! \NR - \! a ) t + \! (N - 1)  (N - 2) }{t^2}} p^{(1)} (t, a) \nonumber \\
  & & \; {- \frac{  2 t^2 - (2  N - 4) t}{t^2}} p^{(2)} (t, a).
\end{eqnarray}
Finally, by defining the function vector
\begin{eqnarray}
\label{equation_up_definition}
\up(t, a) =
 \begin{pmatrix}
p (t, a) &
p^{(1)} (t, a) &
p^{(2)} (t, a)
\end{pmatrix}^{\cal{T}},
\end{eqnarray}
we recast~(\ref{equation_f3}) as the system of differential equations w.r.t.~$ t $
\begin{eqnarray}
\label{equation_up}
\partial_{t}  \up(t, a) = \uP (t, a) \, \up(t, a),
\end{eqnarray}
where the elements of the $ 3 \times 3 $ companion matrix $ \uP (t, a)  $ are:
\begin{eqnarray}
\left[\uP (t, a)\right]_{1, 1} &=&  \left[\uP (t, a)\right]_{1, 3} = 0, \nonumber \\
\left[\uP (t, a)\right]_{2, 1} &=& \left[\uP (t, a)\right]_{2, 2} = 0, \nonumber \\
\left[\uP (t, a)\right]_{1, 2} &=& \left[\uP (t, a)\right]_{2, 3} = 1, \nonumber \\
\left[\uP (t, a)\right]_{3, 1} &=& {\frac{ \!  ( \NR - 2)  t + \! (N - 1) (2 - \! \NR - \! a) }{t^2}}, \nonumber \\
\left[\uP (t, a)\right]_{3, 2}&=& {-\frac{ t^2 + ( 6 - 2 N - \NR - a ) t }{t^2}}, \nonumber \\
&& \quad {-\frac{ (N - 1)  (N - 2)  }{t^2}}, \nonumber \\
 \left[\uP (t, a)\right]_{3, 3} &=& {- \frac{  2 t^2 - (2  N - 4) t}{t^2}}. \nonumber
\end{eqnarray}


Note, however, that the system of differential equations w.r.t.~$t$ in~(\ref{equation_up}) was not deduced for directly solving the original problem. 
Thus, HGM w.r.t.~$ t $ based on~(\ref{equation_up}) shall not be applied to compute the SNR p.d.f.~for a given $ a $ value (e.g., corresponding to a practical $ K $ value) over a relevant SNR range starting at some $ t_0 \neq 0 $. 
Such attempt would require $ p (t_0, a) $, $ p^{(1)} (t_0, a) $, and $ p^{(2)} (t_0, a) $ at the given $ a $, but the infinite series available for them --- shown in~(\ref{equation_gamma1_pdf_final}) and Appendix~\ref{section_pdf_ders_wrt_t} --- can be computed reliably only for small $ a $, as explained in Section~\ref{section_SNR_MGF_PDF_difficult}.

Nevertheless, the system of differential equations w.r.t.~$t$ from~(\ref{equation_up}) shall help indirectly in solving the original problem, because $ \partial_{a} p^{(q)} (t, a) $ can be expressed in terms of $ p^{(q)} (t, a) $, $ q = 0, 1, 2 $, as shown next. 
The ensuing system of differential equations w.r.t.~$ a $ for $ \up (t, a) $ may then be used for HGM-based computation of $ p (t, a) $ at desired values of $ a $. 

\subsection{System of Differential Equations w.r.t.~$ a $ for $ \up (t, a) $}
\label{section_inverse_laplace_2}

In Appendix~\ref{section_differentiation}, Eq.~(\ref{equation_relationship_deriv_wrt_a_s_MGF}), we have deduced the relationship
\begin{eqnarray}
\label{equation_relationship_deriv_wrt_a_s_1}
a\partial_a M(s, a) = \left( s \partial_s - s^2 \partial_s - N {s} \right) M(s, a),
\end{eqnarray}
which, by using rules~(\ref{equation_partials_s_rules_1}) and~(\ref{equation_partials_s_rules_3}), becomes
\begin{eqnarray}
\label{equation_relationship_deriv_wrt_a_s_2}
a\partial_a M(s, a) & = & \left( \partial_s s - 1 - \partial_s s^2 + 2 s - N {s} \right)  M(s, a) \nonumber \\ & = & \left[ - 1 + (- N {s} + \partial_s s + 2 s) - \partial_s s^2 \right]  M(s, a). \nonumber
\end{eqnarray}
Transformation of the above to the $ t $-domain based on~(\ref{equation_diff_integral_rule_1}) and further manipulation yield
\begin{eqnarray}
\label{equation_p_da}
 a \partial_{a} p(t, a) & = & \overbrace{(N - 1) \, p(0+, a)}^{=0, \forall N}-  p(t, a) \nonumber \\ 
& &  + \left(N - t - 2\right) p^{(1)} (t, a) - t \, p^{(2)} (t, a) \nonumber \\
\label{equation_p_da_x}
& = & \uqsimple_1(t, a) \up(t, a),
\end{eqnarray}
where $ \uqsimple_1(t, a) = \begin{pmatrix} -1 & N - t - 2 & -t \end{pmatrix} $.

Now, because $ \partial_{a} p(t, a) $ depends on the entire vector $ \up(t, a) $ defined in~(\ref{equation_up_definition}), we have to also express $ \partial_{a}  p^{(1)} (t, a)  $ and $ \partial_{a}  p^{(2)} (t, a)  $ in terms of $ \up(t, a) $, i.e., we have to express $ \partial_{a} \up (t, a) $ in terms of $ \up(t, a) $.
This is achieved by differentiating~(\ref{equation_p_da_x}) w.r.t.~$ t $ and substituting $ p^{(3)} (t, a) $ from~(\ref{equation_f3}) twice, successively. The procedure yields the following expressions:
\begin{eqnarray}
\label{equation_p_da_1}
&& a \partial_{a}  p^{(1)} (t, a) \nonumber \\
&& = - t \,  p^{(3)} (t, a)  + \left(N - t - 3\right)   p^{(2)} (t, a)  - 2   p^{(1)} (t, a) \nonumber \\
&& = \left(2 - \NR + \frac{2 - 2 N - \NR - a + N \NR + N a}{t} \right) p(t, a) \nonumber\\ 
&& \quad + \left( 4 -2 N - \NR - a +t  + \frac{2 + N^2 - 3 N}{t} \right) p^{(1)} (t, a) \nonumber\\ 
&& \quad +\left( 1 - N + t\right) p^{(2)} (t, a), \\
\label{equation_p_da_2}
&& a \partial_{a}  p^{(2)} (t, a) \nonumber \\
&& = \bigg( -2 + \NR  + \frac{-4 +4 N + 2\NR +a - 2 N \NR -N a }{t} \nonumber \\ 
&& \quad \quad  + \frac{ -4 +6 N + 2\NR + 2 a -3 N \NR - 3 N a }{t^2}  \nonumber \\ 
&& \quad \quad + \frac{N^2 \NR + N^2 a  - 2 N^2}{t^2} \bigg) p(t, a)\nonumber \\ 
&& \quad + \bigg( 3N - 4 + a - t + \frac{ -6 - 3 N^2 + 9 N }{t} \nonumber \\ 
&& \quad \quad + \frac{-4+8N-5N^2 + N^3 }{t^2} \bigg) p^{(1)} (t, a) \nonumber \\ 
&& \quad + \bigg(-1 + 2N -\NR -a -t  \nonumber \\ 
&& \quad \quad+ \frac{ -2 -N^2+3N }{t} \bigg) p^{(2)} (t, a).
\end{eqnarray}
Finally, collecting~(\ref{equation_p_da})--(\ref{equation_p_da_2}) yields for vector $ \up (t, a) $ the system of differential equations w.r.t.~$ a $
\begin{eqnarray}
\label{equation_up_vs_a}
  \partial_{a} \up (t, a) = \frac{1}{a} \, \uQ (t, a) \, \up (t, a),
 \end{eqnarray}
where the first row of the $ 3 \times 3 $ matrix $ \uQ (t, a)  $ is $ \uqsimple_1(t, a)  $ and the remaining elements are as follows:
\begin{eqnarray}
    \left[ \uQ (t, a) \right]_{2,1} & = & 2 - \NR + \frac{2 - 2 N - \NR - a + N \NR + N a}{t}, \nonumber \\
    \left[ \uQ (t, a) \right]_{2,2} & = & 4 -2 N - \NR - a +t  + \frac{2 + N^2 - 3 N}{t}, \nonumber \\
    \left[ \uQ (t, a) \right]_{2,3} & = & 1 - N + t, \nonumber\\
 \left[ \uQ (t, a) \right]_{3,1} & = &  -2 + \NR  + \frac{-4 +4 N + 2\NR }{t} \nonumber \\ 
 && + \frac{a - 2 N \NR -N a }{t} \nonumber \\ 
 &&  + \; \frac{ -4 +6 N + 2\NR + 2 a -3 N \NR - 3 N a}{t^2} \nonumber \\ 
 &&  + \; \frac{N^2 \NR + N^2 a - 2 N^2}{t^2}, \nonumber \\
\left[ \uQ (t, a) \right]_{3,2} & = & 3N - 4 + a - t + \frac{ -6 - 3 N^2 + 9 N }{t}\nonumber \\ &&  + \frac{-4 + 8 N -5 N^2 + N^3 }{t^2}, \nonumber \\
\left[ \uQ (t, a) \right]_{3,3} & = & -1 + 2N -\NR -a -t + \frac{ -2 -N^2+3N }{t}. \nonumber
\end{eqnarray}

Although the above derivations may appear to suggest that the system of differential equations w.r.t.~$ t $ from~(\ref{equation_up}) is solely an instrument for deducing the one w.r.t.~$ a $ from~(\ref{equation_up_vs_a}), we explain next that~(\ref{equation_up}) is as important as~(\ref{equation_up_vs_a}) in the HGM-based computation of the SNR p.d.f..

\subsection{Computation of $ p(t, a) $ by HGM w.r.t.~$ a $, Given $ t $}
\label{section_HGM_wrt_a}

One may attempt to apply HGM w.r.t.~$ a $, i.e., to compute the SNR p.d.f.~by solving~(\ref{equation_up_vs_a}) numerically between some small\footnote{HGM requires $ a_0 \neq 0 $ because $ a $ divides matrix $ \uQ $ in~(\ref{equation_up_vs_a}).} $ a_0 $ and the desired $ a $ (i.e., $ K $), given $ t $ and initial-condition vector $\up(t, a_0) = \left( p (t, a_0) \; p^{(1)} (t, a_0) \; p^{(2)} (t, a_0) \right)^{\cal{T}} $ with elements computed by truncating their infinite series shown in~(\ref{equation_gamma1_pdf_final}) and in Appendix~\ref{section_pdf_ders_wrt_t}.

However, we have found that, for moderate-to-large $ t $, the value of $ p (t, a_0) $ can be too small for sufficiently-accurate numerical representation, which prevents HGM w.r.t.~$a$ from accurately computing $ p(t, a) $ between $ a_0 $ and $ a $.


Nevertheless, interestingly, we can avoid these numerical issues for HGM w.r.t.~$ a $ based on~(\ref{equation_up_vs_a}) by applying it in conjunction with HGM w.r.t.~$ t $ based on~(\ref{equation_up}), as shown next.




\subsection{Computation of $ p(t, a) $ by HGM, for $ a = c \, t $}
\label{section_on_the_line}

In the systems of differential equations obtained in~(\ref{equation_up}) and~(\ref{equation_up_vs_a}), i.e., in
\begin{eqnarray}
\label{equation_diff_eq_t}
\partial_{t}  \up(t, a) & = & \uP (t, a) \, \up(t, a), \\
\label{equation_diff_eq_a}
  \partial_{a} \up (t, a) & = & \frac{1}{a} \, \uQ (t, a) \, \up (t, a),
 \end{eqnarray}
we now make the following changes of variables
\begin{eqnarray}
t & = & c_1 u, \\
a & = & c_2 u.
\end{eqnarray}
Then, the bivariate function vector from~(\ref{equation_up_definition}) becomes the univariate function vector 
\begin{eqnarray}
\label{equation_g}
\up(c_1 u, c_2 u) = 
 \begin{pmatrix}
p (c_1 u, c_2 u) \\
p^{(1)} (c_1 u, c_2 u) \\
p^{(2)} (c_1 u, c_2 u)
\end{pmatrix}= \widetilde{\up} (u).
\end{eqnarray}
Based on the chain rule\cite[Eq.~(1.5.7), p.~7]{NIST_book_10} as well as on~(\ref{equation_diff_eq_t}) and~(\ref{equation_diff_eq_a}), we can write:
\begin{eqnarray}
\frac{d}{d u} \widetilde{\up} (u) & = & \frac{d}{d u} \up(\underbrace{c_1 u}_{t}, \underbrace{c_2 u}_{a}) \nonumber \\ & = & \left[ \partial_{t} \up(t, a) \frac{d t}{d u} + \partial_{a} \up(t, a) \frac{d a}{d u} \right] \bigg|_{{t = c_1 u} \atop {a = c_2 u}}  \nonumber \\
& = & c_1 \, \uP (c_1 u, c_2 u) \, \widetilde{\up} (u) + c_2 \,  \frac{1}{c_2 u} \, \uQ (c_1 u, c_2 u) \, \widetilde{\up} (u) \nonumber \\
& = & c_1 \, \uP (c_1 u, c_2 u) \, \widetilde{\up} (u) + \frac{1}{u} \, \uQ (c_1 u, c_2 u) \, \widetilde{\up} (u).
\end{eqnarray}
Then, for example, $ c_1 = 1 $ and $ c_2 = c $, i.e., $ a = c \, t $, yields the system of differential equations
\begin{eqnarray}
\label{equation_diff_g}
\frac{d}{d u} \widetilde{\up} (u) \! = \left[ \uP (u, c \, u) \! + \! \frac{1}{u} \, \uQ (u, c \, u) \right] \widetilde{\up}(u),
\end{eqnarray}
which helps apply HGM for $ a = c \, t $, i.e., simultaneously w.r.t.~$ a $ and $ t $.

From $ p(t, a) $ computed with the HGM as above we can recover the SNR p.d.f.~$ p_{\gamma_1} (t, a) $ with~(\ref{equation_pdf_relation}). Finally, numerical integration of $ p_{\gamma_1} (t, a) $ yields the outage probability and ergodic capacity based on~(\ref{equation_Po_definition}) and~(\ref{equation_capacity_ergodic_inf_sum}), respectively. \label{pageref_numerical_int}


\section{Numerical Results}
\label{section_pdf_methods_HGMnumerical}

\subsection{Settings and Approach}
\label{section_numres_settings_approach}


The numerical results presented below have been obtained for a MIMO system with $ \NR = 6 $ and $ \NT = 2 $ under Rician--Rayleigh fading with $ K = 7 $~dB and $ \text{AS} = 51^\circ $, i.e., the means of their WINNER II lognormal distributions for the indoor scenario A1\cite{winner_d_1_1_2_v_1_2}, as follows:
\begin{itemize}
\item Monte-Carlo simulation: random samples of the channel matrix $ \uH $ have been generated based on the model~(\ref{equation_channelH}); the ZF SNR for Stream 1 has been computed for each sample by using~(\ref{equation_gammak_perfect_CSI}); a histogram of the samples has yielded the p.d.f.~and c.d.f..
\item Analysis: from expressions~(\ref{equation_gamma1_pdf_Ray}),~(\ref{equation_Po_Rayleigh}) and~(\ref{equation_C_Rayleigh}) and from HGM for differential equation~(\ref{equation_diff_g}). When necessary, we have integrated numerically using the rectangle method\label{pageref_rectangle}.
\end{itemize}

We now outline our HGM-based procedure based on~(\ref{equation_diff_g}) for the computation of the ZF SNR p.d.f.~and ensuing performance measures. Given $ \Gamma_{\text{s}} $ defined in~(\ref{equation_SNR_per_symbol}) and the Rician $ K $-factor, we have computed $ \Gamma_1 $ and $ a $ with~(\ref{equation_Gamma}) and~(\ref{equation_a_first}), respectively. Then, we have computed the Stream-1 ZF SNR p.d.f.~over the SNR range with the following steps:
\begin{enumerate}
\item Compute accurately the initial condition\footnote{Note that both $ t $ and $ a $ are substituted with $ u_0 $ in this step, for simplicity.} $ \widetilde{\up} (u_0) = \left( p (u_0, u_0) \; p^{(1)} (u_0, u_0) \; p^{(2)} (u_0, u_0) \right)^{\cal{T}} $ for a sufficiently-small $ u_0 $, using the infinite series for $ p^{(q)} (t, a)  $ derived in Appendix~\ref{section_pdf_ders_wrt_t}.
\item Sample the SNR range of interest $ [ u_1, \, u_M ] $ as $ u_1, \, u_2, \, \cdots, \, u_M $.
\item For each sample $ u = u_m $, $ m = 1, \, 2, \, \cdots \, , M $: first, set $ c = a/u $; then, apply HGM to solve~(\ref{equation_diff_g}) from $ u_0 $ to $ u $; finally, save $ p (u, c u)  $, i.e., $ p(t, a) $ on the line $ a = c \, t $.
\item Recover the ZF SNR p.d.f.~$ p_{\gamma_1} (t, a)  $ from $ p(t, a) $ by~(\ref{equation_pdf_relation}).
\end{enumerate}
This approach avoids using the infinite series for $ p (t, a) $, $ p^{(1)} (t, a) $, and $ p^{(2)} (t, a) $ at large $ a $ or large $ t $, and, thus, avoids the  numerical issues described in Sections~\ref{section_SNR_MGF_PDF_difficult} and~\ref{section_HGM_wrt_a}.

Finally, we have computed the outage probability and ergodic capacity by numerically integrating, according to~(\ref{equation_Po_definition}) and~(\ref{equation_capacity_ergodic_inf_sum}), respectively, the p.d.f.~produced by HGM as described above.

\subsection{Description of Results}
\label{section_results_description}

\begin{figure}[t]
\begin{center}
\includegraphics[width=3.5in]
{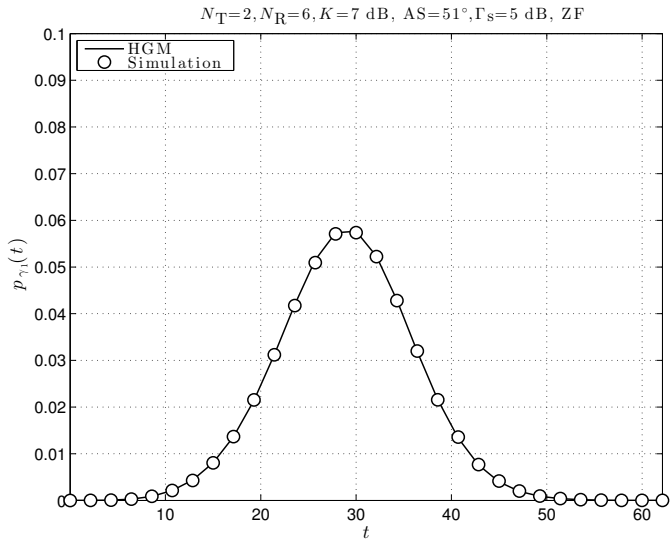}
\caption{Stream-1 SNR p.d.f.~computed by the HGM, based on~(\ref{equation_diff_g}), and by Monte-Carlo simulation, for $ \NR = 6 $, $ \NT = 2 $, and Rician--Rayleigh fading with $ K = 7 $~dB, $ \text{AS} = 51^\circ $.}
\label{figure_pdf_vs_t_Method_4_NT_2_NR_6_K_7}
\end{center}
\end{figure}

\begin{figure}[t]
\begin{center}
\includegraphics[width=3.5in]
{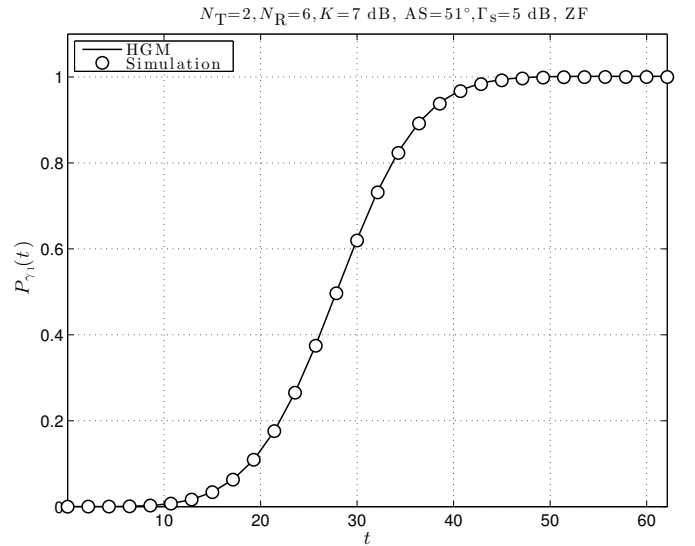}
\caption{Stream-1 SNR c.d.f.~computed by numerical integration of the p.d.f.~produced by HGM, based on~(\ref{equation_diff_g}), and by Monte-Carlo simulation, for $ \NR = 6 $, $ \NT = 2 $, and Rician--Rayleigh fading with $ K = 7 $~dB, $ \text{AS} = 51^\circ $.}
\label{figure_cdf_vs_t_Method_4_NT_2_NR_6_K_7}
\end{center}
\end{figure}

Figs.~\ref{figure_pdf_vs_t_Method_4_NT_2_NR_6_K_7} and~\ref{figure_cdf_vs_t_Method_4_NT_2_NR_6_K_7} depict, respectively, the SNR p.d.f.~and c.d.f.~computed with the HGM as above, and by simulation.
The HGM is successful, i.e., the resulting p.d.f.~and c.d.f.~agree with the simulation, and the c.d.f.~shown in~Fig~\ref{figure_cdf_vs_t_Method_4_NT_2_NR_6_K_7} (from numerically integrating the p.d.f.~produced by HGM) goes to $ 1 $ for increasing $ t $.
Recall that in\cite{siriteanu_twc_13}\cite{siriteanu_ausctw_14} we had been able to accurately compute $ p_{\gamma_1} (t) $ based on its infinite series~(\ref{equation_gamma1_pdf_final}) only up to the unrealistically-small $ K $ value of $ 1.5 $~dB. 
This has also been illustrated herein in Fig.~\ref{figure_ZF_SNR_pdf_Rayleigh_Rice_Fixed_AS_K_70_A1_NT2_NR6}, where the series-based computation breaks down for $ K = 7 $~dB. 
Consequently, Figs.~\ref{figure_pdf_vs_t_Method_4_NT_2_NR_6_K_7} and~\ref{figure_cdf_vs_t_Method_4_NT_2_NR_6_K_7} do not attempt to plot series results.

\begin{figure}[t]
\begin{center}
\includegraphics[width=3.5in]
{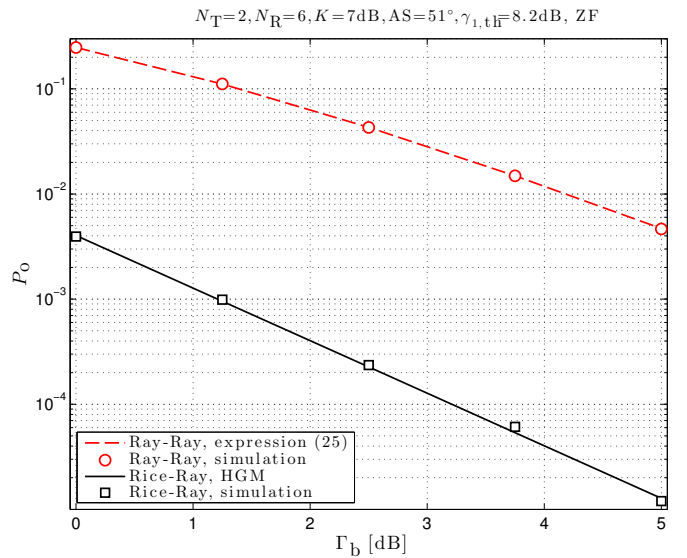}
\caption{Stream-1 outage probability for $ \NR = 6 $, $ \NT = 2 $, and fading parameters $ K = 7 $~dB, $ \text{AS} = 51^\circ $.
For Rayleigh-only fading: from expression~(\ref{equation_Po_Rayleigh}) and from simulation.
For Rician--Rayleigh fading: from numerical integration according to~(\ref{equation_Po_definition}) of the SNR p.d.f.~computed with the HGM, based on~(\ref{equation_diff_g}), and from simulation.}
\label{figure_ZF_SNR_Po_Rayleigh_Rice_Fixed_AS_K_7_A1_NT2_NR6}
\end{center}
\end{figure}

\begin{figure}[t]
\begin{center}
\includegraphics[width=3.5in]
{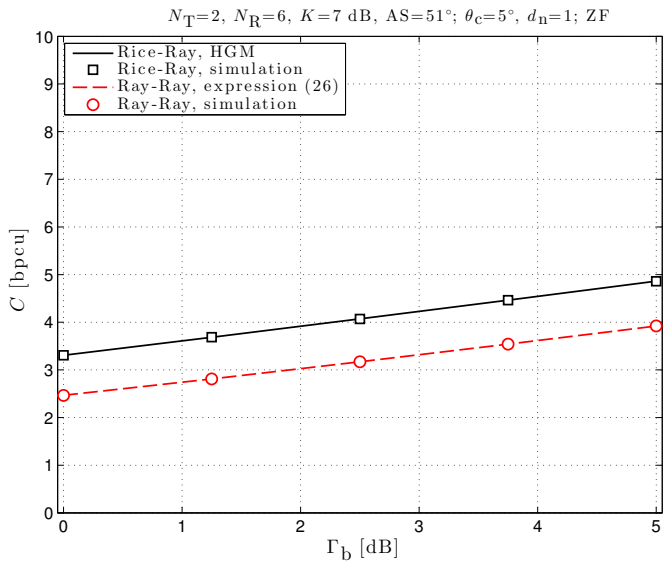}
\caption{Stream-1 ergodic capacity for $ \NR = 6 $, $ \NT = 2 $, and fading parameters $ K = 7 $~dB, $ \text{AS} = 51^\circ $.
For Rayleigh-only fading: from expression~(\ref{equation_C_Rayleigh}) and from simulation.
For Rician--Rayleigh fading: from numerical integration according to~(\ref{equation_capacity_ergodic_inf_sum}) of the SNR p.d.f.~computed by HGM, based on~(\ref{equation_diff_g}), and from simulation.}
\label{figure_ZF_SNR_C_Rayleigh_Rice_Fixed_AS_K_7_A1_NT2_NR6}
\end{center}
\end{figure}

Finally, Figs.~\ref{figure_ZF_SNR_Po_Rayleigh_Rice_Fixed_AS_K_7_A1_NT2_NR6} and~\ref{figure_ZF_SNR_C_Rayleigh_Rice_Fixed_AS_K_7_A1_NT2_NR6} depict, respectively, the outage probability and ergodic capacity (in bpcu) with respect to $ \Gamma_{\text{b}} $ defined in~(\ref{equation_SNR_per_bit}).
For Rayleigh-only fading we have used the integral expressions~(\ref{equation_Po_Rayleigh}) and~(\ref{equation_C_Rayleigh}), respectively.
For Rician--Rayleigh fading we have integrated numerically\label{pageref_numerical_integration} according to~(\ref{equation_Po_definition}) and~(\ref{equation_capacity_ergodic_inf_sum}), respectively, the SNR p.d.f.~produced by HGM as shown above.
The HGM and simulation results agree closely\footnote{Unshown results have revealed that HGM yields accurate results even for $ K $ as high as $ 15 $~dB, and also for other combinations of $ \NT $ and $ \NR $, without a noticeable increase in computation time.}. 
On the other hand, results from $ P_{\text{o}} $ and $ C $ infinite series\cite[Eqs.~(69), (71)]{siriteanu_twc_13} could not be shown because their computation breaks down.

\subsection{HGM Complexity}
\label{section_HGM_complexity}

Our HGM-based computation solves~(\ref{equation_diff_g}) with the iterative Runge--Kutta method\cite[Section~3.7]{NIST_book_10}\cite{sei_sc_13}, for tolerance level $ \epsilon = 10^{-15} $ (i.e., $ 15 $ digits of accuracy), with the \texttt{MATLAB ode} function.
Note that the Runge-Kutta method is available in most numerical tools, and its complexity is polynomial in the number of digits of accuracy\cite{ilie_jc_08}.
Thus, we have found that the duration of the HGM-based computation is reasonable\footnote{Computation of the p.d.f.~at $ 30 $ samples of $ t $, as shown in Fig.~\ref{figure_pdf_vs_t_Method_4_NT_2_NR_6_K_7} requires about $ 60 $ seconds. Then, outage probability computation for one $ \Gamma_{\text{b}} $ value takes about $ 70 $~s. Finally, ergodic capacity computation for one $ \Gamma_{\text{b}} $ value takes about $ 140 $~s.}. Finally, HGM-based computation duration and success are robust to the value of $ K $, unlike infinite-series-based computation\cite{siriteanu_twc_13}\cite{siriteanu_ausctw_14}\cite{muller_nm_01}.

\section{Future Work: an HGM-Based Framework for MIMO Evaluation}
\label{section_HGM_MIMO_Applications}

\subsection{HGM-Based Evaluation of MIMO under General Fading}
\label{section_HGM_Other_Fading}
For many fading types (e.g., Rayleigh, Rician, Nakagami, lognormal) and for many multiantenna transmission techniques, previous work, e.g.\cite{simon_alouini_book_00}, showed expressions for the SNR m.g.f.~that involve special (e.g., hypergeometric, Bessel) functions of scalar argument, which have typically been written as infinite-series, but are also holonomic. 
On the other hand,\cite{ropokis_tcom_10} analyzed MIMO performance under general fading by purposely writing as infinite series (ensuing from expansions around $ 0 $) the SNR m.g.f..
We have been applying a similar approach in\cite{siriteanu_twc_14_NRX2} for $ \NR \times 2 $ MIMO ZF under full-Rician fading. 
The deduced SNR p.d.f.~expression contains multiple infinite series whose computation by truncation has once again been found accurate only for unrealistically-small $ K $ values.
Therefore, we are currently attempting to apply instead the HGM-based approach: deduce and solve relevant differential equations.

Furthermore, hypergeometric functions also of matrix argument have often occurred in MIMO analyses due to statistical assumptions about the channel matrix\cite{koev_mc_06}\cite{grant_tcom_05}.
For example, the c.d.f.~and m.g.f.~of the dominant eigenvalue of a complex-valued central-Wishart-distributed matrix have been expressed in terms of $ {_1\! F_1}(a; c; \uR) $ and $ {_2\! F_1}(a; c; \uR) $ in\cite[Eqs.~(34), (42)]{grant_tcom_05}, respectively. 
Thus, for MIMO beamforming under Rayleigh-fading channel, the average error probability and outage probability have been expressed in terms of $ {_1\! F_1}(a; c; \uR) $ and $ {_2\! F_1}(a; c; \uR) $ in\cite[Eqs.~(30), (22)]{grant_tcom_05}, respectively. 
Unfortunately, the well-known infinite series for these functions (and the zonal polynomials involved)\cite[Eq.~(1.1)]{muirhead_ams_70}\cite[Eq.~(1.1)]{koev_mc_06}\cite[Eq.~(61)]{grant_tcom_05} are difficult to compute\cite{koev_mc_06}\cite{grant_tcom_05}.
Nevertheless, such functions also satisfy differential equations\cite{muirhead_ams_70}\cite{chikuse_aism_76}.
Those for $  {_1\! F_1}(a; c; \uR) $ from\cite[Eq.~(5.1)]{muirhead_ams_70} have recently been exploited for the accurate HGM-based computation of the c.d.f.~of the dominant eigenvalue of a real-valued central-Wishart-distributed matrix in\cite{hashiguchi_jma_2013}.
We shall pursue this approach for complex-valued matrices, in order to evaluate MIMO beamforming performance.

\subsection{Computer-Algebra-Based Deduction of Equations}
\label{section_automation}

In the current paper, HGM has been applied for solving differential equations for the ZF SNR m.g.f.~and p.d.f.~that we deduced manually. 
Computer-algebra tools that can help automate the deduction of differential equations for holonomic functions have recently become available\cite[p.~171]{kauers_book_11}\cite[Ch.~7]{hibi_book_13}\cite{koutschan_thesis_09}\cite{koutschan_tr_10b}. 
Thus, in other recent work\cite{siriteanu_icc_15}, we have investigated employing and enhancing such tools to automatically derive differential equations not only for the ZF SNR m.g.f.~and p.d.f., but also for the ZF outage probability and ergodic capacity. 
We shall continue investigating whether such automated tools can help apply HGM to MIMO performance evaluation under general fading, by deducing differential equations --- instead of infinite series as in\cite{simon_alouini_book_00}\cite{ropokis_tcom_10}\cite{siriteanu_twc_14_NRX2}.

The outcome of this future work is envisioned to become a framework for the automated analysis and HGM-based evaluation of MIMO performance under general fading.

\section{Summary and Conclusions}
For MIMO ZF under Rician--Rayleigh fading, this paper has demonstrated that performance-measure expressions can be evaluated accurately, by using the HGM, at (realistic) Rician $ K $-factor values that render unusable the conventional method of truncating infinite series.
For the SNR m.g.f., which has been known in terms of the confluent hypergeometric function, which is a holonomic function, we have deduced the satisfied differential equations.
They have yielded the differential equations satisfied by the SNR p.d.f., which, in turn, have helped compute the p.d.f.~accurately using the HGM at values of $ K $ relevant according to WINNER II (e.g., $ K = 7 $~dB).
Finally, numerical integration of the SNR p.d.f.~obtained by the HGM has yielded for the MIMO ZF outage probability and ergodic capacity close agreement with simulations.

Future work shall attempt to extend the results of this paper into an automated HGM-based analysis and evaluation framework that promises to accurately characterize MIMO performance for realistic fading-parameter values.

\begin{appendices}

\section{Differential Equation w.r.t.~$ s $ for $ M(s, a)  $}
\label{section_derivation_SNR_MGF_ODEs}

First, substituting $ \sigma $ with $ \frac{a s}{1 - s} $ in the differential equation for $ {_1\! F_1} (N; \NR; \sigma) $ from~(\ref{equation_1F1_diff_eq}) yields
\begin{eqnarray}
\label{equation_1F1_diff_eq_sigma_1}
&& \frac{a s}{1 - s}  {_1\! F_1^{(2)}} \left(N; \NR; \frac{a s}{1 - s} \right)   \nonumber \\ 
&& \quad +   \left( \NR - \frac{a s}{1 - s} \right)    {_1\! F_1^{(1)}} \left( N; \NR; \frac{a s}{1 - s} \right)    \nonumber \\ 
&&  \quad -   N  {_1\! F_1} \left( N; \NR; \frac{a s}{1 - s} \right)   = 0.
\end{eqnarray}
Then, from~(\ref{equation_gamma1_mgf_final_scaled_x}), we have
\begin{eqnarray}
\label{equation_gamma1_mgf_final_scaled_x_copy}
M(s, a) = \frac{1}{\left( 1 - s \right)^N} {_1\! F_1} \left(N; \NR;  \frac{a s}{1 - s} \right), 
\end{eqnarray}
which yields
\begin{eqnarray}
\label{equation_1F1_and_derivs_0_1}
{_1\! F_1} \left(N; \NR;  \frac{a s}{1 - s} \right) = \left( 1 - s \right)^{N} M(s, a).
\end{eqnarray}
%
Differentiating~(\ref{equation_gamma1_mgf_final_scaled_x_copy}) w.r.t.~$ s $ yields:
\begin{eqnarray}
\label{equation_gamma1_mgf_final_scaled_diff_wrt_s_2}
\partial_s M(s, a) & = & \frac{N}{\left( 1 - s \right)^{N+1}} {_1\! F_1} \left(N; \NR;  \frac{a s}{1 - s} \right) \nonumber \\ 
&& + \frac{a}{\left( 1 - s \right)^{N+2}} {_1\! F_1^{(1)}} \left(N; \NR;  \frac{a s}{1 - s} \right)
\end{eqnarray}
By first substituting~(\ref{equation_1F1_and_derivs_0_1}) into~(\ref{equation_gamma1_mgf_final_scaled_diff_wrt_s_2}) and then by differentiating the result w.r.t.~$ s $ we obtain
\begin{eqnarray}
\label{equation_relationship_MGF_deriv_wrt_s_a_1}
\partial_s M(s, a) & = & \frac{N}{\left( 1 - s \right)} M(s, a)  \nonumber \\ 
&& + \frac{a}{\left( 1 - s \right)^{N+2}} {_1\! F_1^{(1)}} \left(N; \NR;  \frac{a s}{1 - s} \right) \\
\label{equation_relationship_MGF_deriv_wrt_s_a_2}
\partial^2_s M(s, a) & = & \frac{N}{\left( 1 - s \right)^2} M(s, a) + \frac{N}{\left( 1 - s \right)} \partial_s M(s, a) \nonumber \\ && + \frac{a (N + 2)}{\left( 1 - s \right)^{N+3}} {_1\! F_1^{(1)}} \left(N; \NR;  \frac{a s}{1 - s} \right)    \nonumber \\ 
&&  +   \frac{a^2}{\left( 1 - s \right)^{N+4}} {_1\! F_1^{(2)}} \left(N; \NR;  \frac{a s}{1 - s} \right)
\end{eqnarray}
which yield, respectively:
\begin{eqnarray}
\label{equation_relationship_1F1_MGF_1}
&&{_1\! F_1^{(1)}} \left(N; \NR;  \frac{a s}{1 - s} \right)  \nonumber \\ 
&& \quad = \frac{\left( 1 - s \right)^{N+2}}{a} \left[ \partial_s - \frac{N}{\left( 1 - s \right)} \right]  M(s, a), \\
\label{equation_relationship_1F1_MGF_2}
&&{_1\! F_1^{(2)}} \left(N; \NR;  \frac{a s}{1 - s} \right)  \nonumber \\ 
&& \quad = \frac{\left( 1 - s \right)^{N+4}}{a^2} \bigg[ \partial^2_s M(s, a) - \frac{N}{\left( 1 - s \right)^2} M(s, a) \nonumber \\ 
& & \quad \quad - \frac{N}{\left( 1 - s \right)} \partial_s M(s, a) \nonumber \\ 
& & \quad \quad - \frac{a (N + 2)}{\left( 1 - s \right)^{N+3}} {_1\! F_1^{(1)}} \left(N; \NR;  \frac{a s}{1 - s} \right) \bigg].
\end{eqnarray}
Substituting~(\ref{equation_relationship_1F1_MGF_1}) into~(\ref{equation_relationship_1F1_MGF_2}) yields:
\begin{eqnarray}
\label{equation_relationship_1F1_MGF_2_1}
{_1\! F_1^{(2)}} \left(N; \NR;  \frac{a s}{1 - s} \right) & = & \frac{\left( 1 - s \right)^{N+4}}{a^2} \bigg[ \partial^2_s - \frac{2 ( N + 1 )}{\left( 1 - s \right)} \partial_s  \nonumber \\ 
&& \quad+ \frac{N (N + 1)}{\left( 1 - s \right)^{2}} \bigg]  M(s, a).
\end{eqnarray}

Finally, substituting~(\ref{equation_1F1_and_derivs_0_1}),~(\ref{equation_relationship_1F1_MGF_1}), and~(\ref{equation_relationship_1F1_MGF_2_1}) into the differential equation~(\ref{equation_1F1_diff_eq_sigma_1}), and further manipulation,  yield the following differential equation w.r.t.~$ s $ for $ M(s, a) $
\begin{eqnarray}
\label{equation_MGF_diff_eq_wrt_s_appendix}
&& \bigg( s (1 - s)^2 \partial_s^2 - \big[ 2 (N + 1) s (1 - s) - (1 - s) \NR + a s \big] \partial_s \nonumber \\ 
&& \quad+ N \big[ (N + 1) s -\NR - a \big]  \bigg)  M(s, a)   =  0,
\end{eqnarray} 
which appears in the main text in~(\ref{equation_MGF_diff_eq_wrt_s_final_main}).

\section{Initial Condition $ p_{\gamma_1}(0+, a) $}
\label{section_initial_condition}

For the special case with $ N = 1 $, i.e., for $ \NR = \NT $,~(\ref{equation_gamma1_pdf_final}) becomes
\begin{eqnarray}
\label{equation_gamma1_pdf_final_N_1}
p_{\gamma_1} (t, a) = \frac{ e^{-t/\Gamma_1}}{\Gamma_1} \sum_{n = 0}^{\infty} A_n(a) \sum_{m = 0}^{n} {n \choose m}  
\frac{(-1)^{m} t^{ n - m }  }{(n - m)! \,  \Gamma_1^{n - m}}, \nonumber
\end{eqnarray}
which yields
\begin{eqnarray}
\label{equation_gamma1_pdf_final_N_1_0_plus}
\lim_{t \rightarrow 0, t > 0} p_{\gamma_1} (t, a) & = & p_{\gamma_1} (0+, a) = \frac{ 1}{\Gamma_1} \sum_{n = 0}^{\infty} A_n(a) (-1)^{n} \nonumber \\ 
&=& \frac{ 1}{\Gamma_1} \sum_{n = 0}^{\infty} { \frac{\left( N \right)_n}{\left( \NR \right)_n} \frac{(-a)^n}{n!} }.
\end{eqnarray}
Thus,~(\ref{equation_gamma1_pdf_final_N_1_0_plus}),~(\ref{equation_1F1_series}), and~(\ref{equation_gamma1_pdf_final}) yield
\begin{eqnarray}
\label{equation_p__gamma1_0_plus}
p_{\gamma_1}(0+, a) = 
\begin{cases}
\frac{ 1}{\Gamma_1}  {_1\! F_1}(N; \NR; -a), & N = 1, \\
0, & N > 1,
\end{cases}
\end{eqnarray}
which is used in the main text to deduce~(\ref{equation_p_0_plus}).

\section{Infinite Series for Derivatives of $ p(t, a) $~w.r.t.~$ t $}
\label{section_pdf_ders_wrt_t}

Based on~(\ref{equation_gamma1_pdf_final}) and~(\ref{equation_pdf_relation}), let us define the function
\begin{eqnarray}
\label{equation_f_0}
f(t, a) = p (t, a) e^t & = & {\sum_{n = 0}^{\infty}   A_n(a)   \sum_{m = 0}^{n}   {n \choose m} }\nonumber \\ 
&& \times
\frac{ (-1)^{m} t^{N + n - m - 1} }{(N + n - m - 1)! },
\end{eqnarray}
whose first two derivatives are given by
\begin{eqnarray}
\label{equation_f_1}
f^{(1)}(t, a) & = & p^{(1)} (t, a) e^t + p (t, a) e^t, \nonumber \\
\label{equation_f_2}
f^{(2)}(t, a) & = & p^{(2)} (t, a) e^t + 2 p^{(1)} (t, a) e^t + p (t, a) e^t. \nonumber
\end{eqnarray}
The above yield
\begin{eqnarray}
\label{equation_f_0_inv}
p(t, a) = f (t, a) e^{-t}, 
\end{eqnarray}
as well as 
\begin{eqnarray}
\label{equation_f_1_inv}
p^{(1)} (t, a) & = &   \left[ f^{(1)}(t, a) - f (t, a) \right] e^{-t},  \\
\label{equation_f_2_inv}
p^{(2)} (t, a) & = &   \left[ f^{(2)}(t, a) - 2 f^{(1)}(t, a)\! + \! f (t, a) \right] \!  e^{-t}
\end{eqnarray}
which are the only derivatives of $ p(t,a) $ required for~(\ref{equation_up_definition}).

Now, if we rewrite $ f(t, a) $ from~(\ref{equation_f_0}) further as
\begin{eqnarray}
\label{equation_f_0_123}
f(t, a) & = & t^{N - 1} \underbrace{{\sum_{n = 0}^{\infty} A_n(a) \sum_{m = 0}^{n} {n \choose m} }\frac{(-1)^{m}t^{n - m} }{(N - 1 + n - m)!} }_{g(t,a)} \nonumber \\
& = & t^{N - 1} g(t,a),
\end{eqnarray}
then its $ q $th partial derivative w.r.t.~$ t $ is\footnote{Note that~(\ref{equation_f_derivative_2}) follows from~(\ref{equation_f_derivative_1}) only for $  k \le N - 1 $.}, based on Leibniz's formula\cite[Eq.~(1.4.12), p.~5]{NIST_book_10}:
\begin{eqnarray}
\label{equation_f_derivative_1}
f^{(q)}(t, a) & = &     {\partial_t^q \left[ t^{N - 1} g(t,a) \right]} \nonumber \\
& = &  \sum_{k = 0}^{q} {q \choose k} \left[ t^{N - 1} \right]^{(k)} g^{(q-k)}(t, a) \\
\label{equation_f_derivative_2}
& = &     \sum_{k = 0}^{q} {q \choose k} \frac{(N - 1)! t^{N - 1 - k} g^{(q-k)}(t, a)}{(N - 1 - k)!}.
\end{eqnarray}
If we rewrite $ g(t,a) $ from~(\ref{equation_f_0_123}) as
\begin{eqnarray}
g(t, a) = {\sum_{n = 0}^{\infty} A_n(a) \sum_{r = 0}^{n} {n \choose n - r} }\frac{(-1)^{n-r}t^{r} }{(N - 1 + r)! }, \nonumber
\end{eqnarray} 
then its partial derivative of order $ q \ge 1 $ w.r.t.~$t$ is given by
\begin{eqnarray}
\label{equation_g_derivatives}
g^{(q)}(t, a) & = & {\sum_{n = q}^{\infty} A_n(a) \sum_{r = q}^{n} {n \choose n - r} } \nonumber \\
&  & \quad \times \frac{(-1)^{n-r}}{(N - 1 + r)! }\frac{r!}{(r-q)!}t^{r-q},
\end{eqnarray}
which, along with~(\ref{equation_f_derivative_2}), yields $ f^{(q)}(t, a) $. 
Finally, substituting into~(\ref{equation_f_1_inv}) and~(\ref{equation_f_2_inv}) yields expressions for $ p^{(1)} (t, a) $ and $ p^{(2)} (t, a) $, respectively.

However, because~(\ref{equation_f_derivative_2}) follows from~(\ref{equation_f_derivative_1}) only for $  k \le N - 1 $, and because $ k $ goes from $ 0 $ to $ q $, it is required that $ N - 1 \ge q $.
Then, because~(\ref{equation_up_definition}) requires $ f^{(q)}(t, a) $ for $ q $ as high as $ 2 $, $ f^{(q)}(t, a)  $ can be written as in~(\ref{equation_f_derivative_2}) only if $ N \ge 3 $.
Table~\ref{table_derivatives_N_1_2} characterizes the remaining cases.

\begin{center}
\begin{table}[t]
\caption{Derivatives of $ f(t,a) $ for $ N = 1, 2 $}
\renewcommand{\arraystretch}{1.35}
\label{table_derivatives_N_1_2}
\centering
\begin{tabular}{c|c|c}
  \hline
    & $ N = 1 $ &  $ N = 2 $\\
   \hline 
   $ f(t, a) $ & $ g(t, a) $ & $ t g(t, a) $ \\
   \hline 
   $ f^{(1)}(t, a) $ & $ g^{(1)}(t, a) $ & $ g(t, a)  + t g^{(1)}(t, a) $ \\
   \hline 
   $ f^{(2)}(t, a) $ & $ g^{(2)}(t, a) $ & $ 2 g^{(1)}(t, a) + t g^{(2)}(t, a) $ \\
  \hline 
\end{tabular}
\end{table}
\end{center}


\section{Relationship Between Derivatives of $ M(s, a) $ w.r.t.~$s$ and $a$}
\label{section_differentiation}

Differentiating~(\ref{equation_gamma1_mgf_final_scaled_x_copy}) w.r.t.~$ a $ yields
\begin{eqnarray}
\label{equation_gamma1_mgf_final_scaled_diff_wrt_a_1}
\partial_a M(s, a) = \frac{s}{\left( 1 - s \right)^{N+1}} {_1\! F_1^{(1)}} \left(N; \NR;  \frac{a s}{1 - s} \right),
\end{eqnarray}
so that
\begin{eqnarray}
\label{equation_1F1_and_derivs_1}
{_1\! F_1^{(1)}} \left(N; \NR;  \frac{a s}{1 - s} \right) = \frac{\left( 1 - s \right)^{N+1}}{s} \partial_a  M(s, a).
\end{eqnarray}
Now, by substituting~(\ref{equation_1F1_and_derivs_0_1}) and~(\ref{equation_1F1_and_derivs_1}) into~(\ref{equation_gamma1_mgf_final_scaled_diff_wrt_s_2}), and by further manipulation, we obtain
\begin{eqnarray}
\label{equation_relationship_deriv_wrt_a_s_MGF}
a \partial_a M(s, a) = {s \left( 1 - s \right)} \partial_s M(s, a) - N {s} M(s, a),
\end{eqnarray}
which appears in the main text in~(\ref{equation_relationship_deriv_wrt_a_s_1}).

%

\end{appendices}

\section*{Acknowledgments}
\addcontentsline{toc}{section}{Acknowledgment}
The first version of this paper was prepared when the first author was with the Graduate School of Information Science and Technology, University of Tokyo, supported by Japan Science and Technology Agency. These institutions also supported the publication of this paper. 
The second (accepted) version was prepared after he joined the Graduate School of Information Science and Technology, Osaka University.

Akimichi Takemura acknowledges the support of the Japan Society for the Promotion of Science (JSPS) grant-in-aid for scientific research No. 25220001 and the support of Japan Science and Technology Agency.

Hyundong Shin acknowledges the support of the National Research Foundation of Korea (NRF) grants No. 2009-0083495 and No. 2013-R1A1A2-019963, funded by the Ministry of Science, ICT \& Future Planning (MSIP).

Christoph Koutschan acknowledges the support of the Austrian Science Fund (FWF): W1214.

\footnotesize


\begin{IEEEbiography}[{\includegraphics[width=1in,height=1.25in,clip,keepaspectratio]{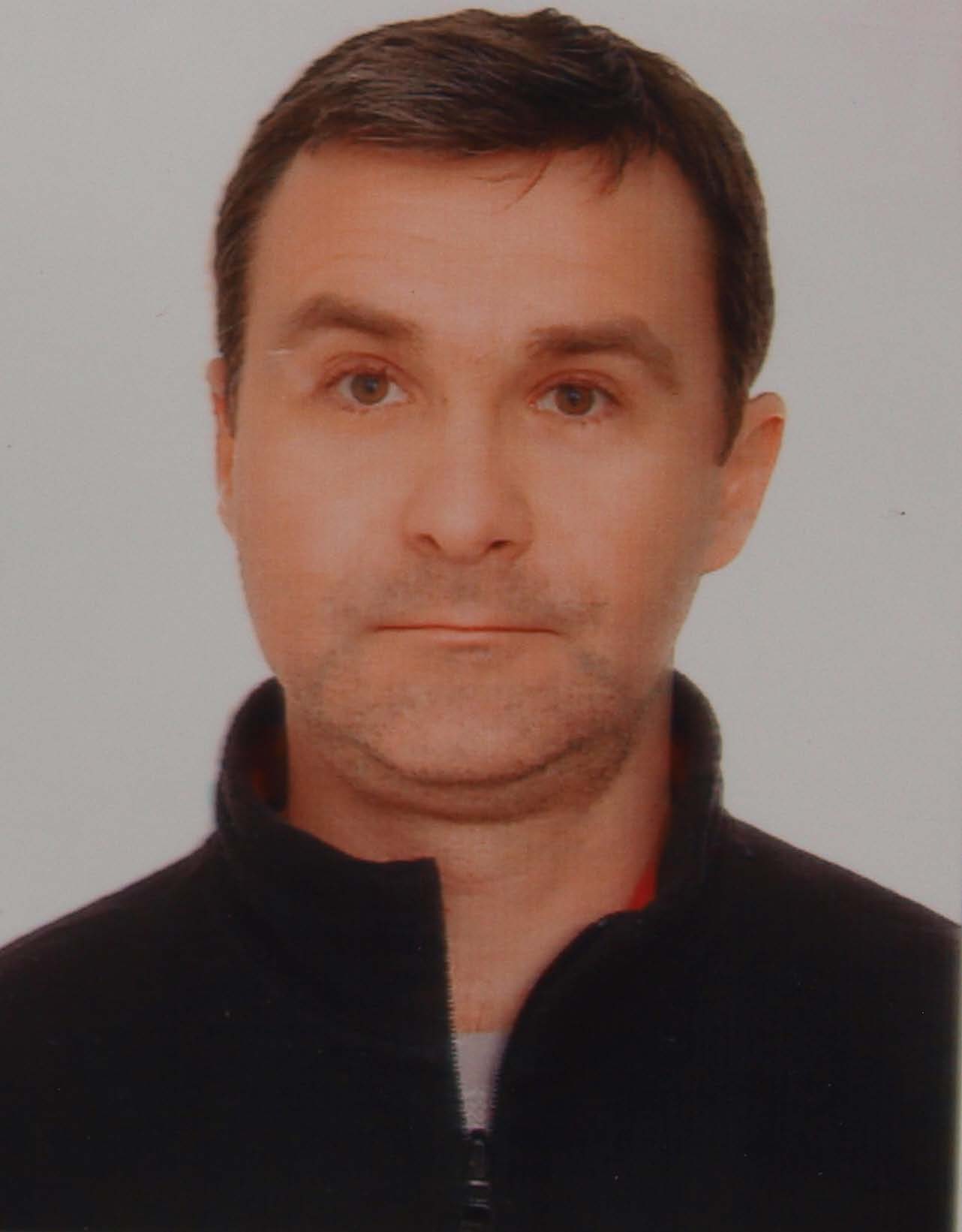}}]{Constantin (Costi) Siriteanu}
was born in
Sibiu, Romania. He received the Bachelor
and Master degrees in  Control Systems
from ``Gheorghe Asachi" Technical University, Iasi,
Romania, in 1995 and 1996, respectively, and the
Ph.D. degree in Electrical and Computer Engineering from Queen's University, Canada, in 2006.
His Ph.D. thesis was on the performance--complexity tradeoff for smart
antennas.
Between September 2006 and March 2014 he worked as Researcher and Assistant Professor in Korea (Seoul National University, Kyung Hee Univesity, Hanyang University), Canada (Queen's University), and Japan (Hokkaido University, University of Tokyo).
Since April 2014, he is a CAREN Specially-Appointed Assistant Professor with the Graduate School of Information Science and Technology, Osaka University.
His research interests have been in developing multivariate statistics concepts
that help analyze and evaluate the performance of multiple-input/multiple-output (MIMO) wireless
communications systems under realistic statistical assumptions about channel fading.
Recently, Constantin has been working on applications of computer algebra to the deduction of implicit representations of MIMO performance measures (i.e., as solutions of differential equations).
\end{IEEEbiography}

\begin{IEEEbiography}[{\includegraphics[width=1in,height=1.25in,clip,keepaspectratio]{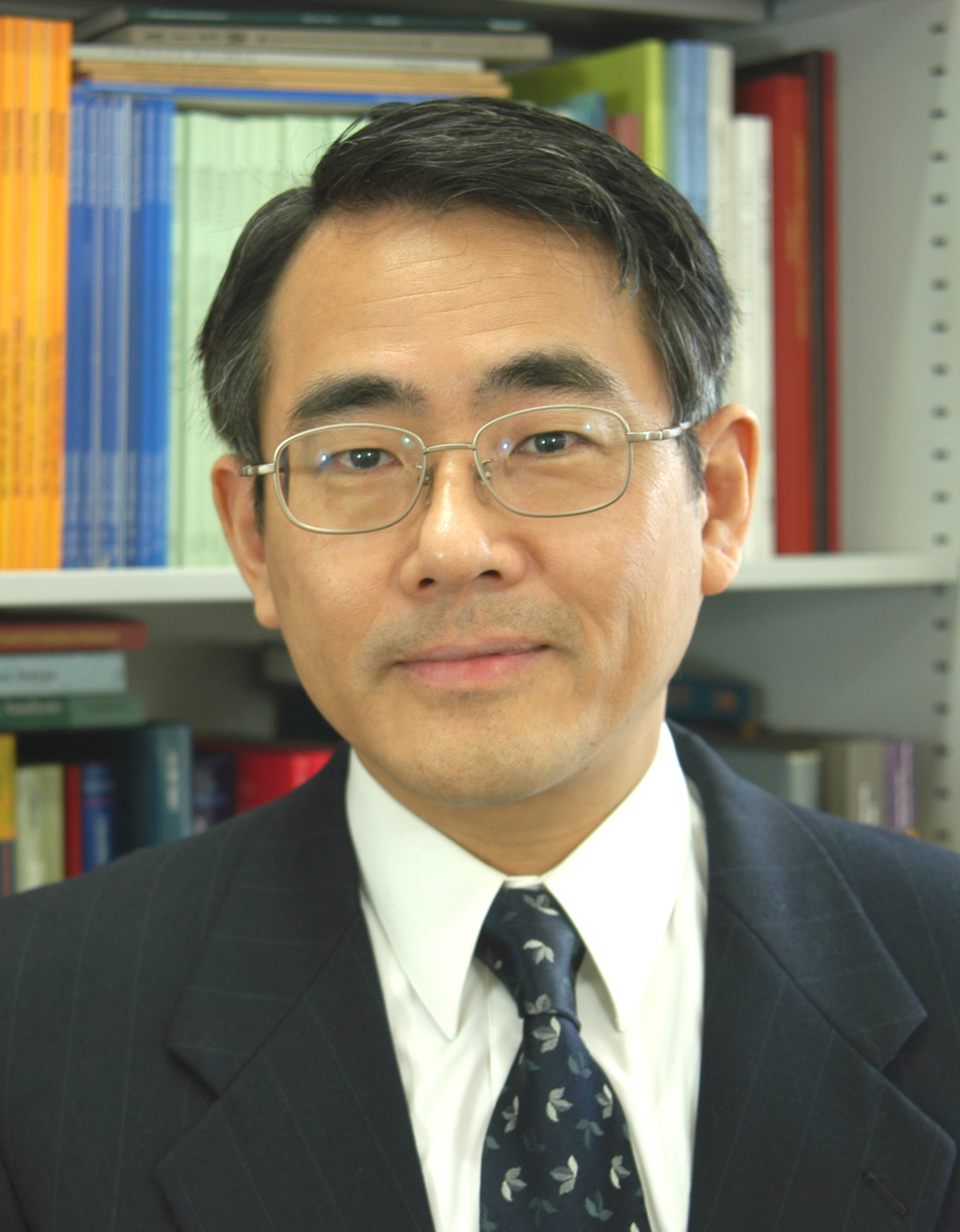}}]{Akimichi Takemura}
received the Bachelor of Arts degree in Economics in 1976 and
the Master of Arts degree in Statistics in 1978 from University of Tokyo, and
the Ph.D. degree in Statistics in 1982 from Stanford University.
He was an acting Assistant Professor at the Department of Statistics, Stanford University
from September 1992 to June 1983, and a
visiting Assistant Professor at the Department of Statistics, Purdue
University from  September 1983 to May 1984.
In June 1984 he has joined University of Tokyo, where he has been a Professor of Statistics with the Department of Mathematical Informatics since April 2001.
He has served as President of Japan Statistical Society from January 2011 to June 2013.
He has been working on multivariate distribution theory in statistics.
Currently his main area of research is algebraic statistics.  He also works on game-theoretic probability,
which is a new approach to probability theory.
\end{IEEEbiography}

\begin{IEEEbiography}[{\includegraphics[width=1in,height=1.25in,clip,keepaspectratio]{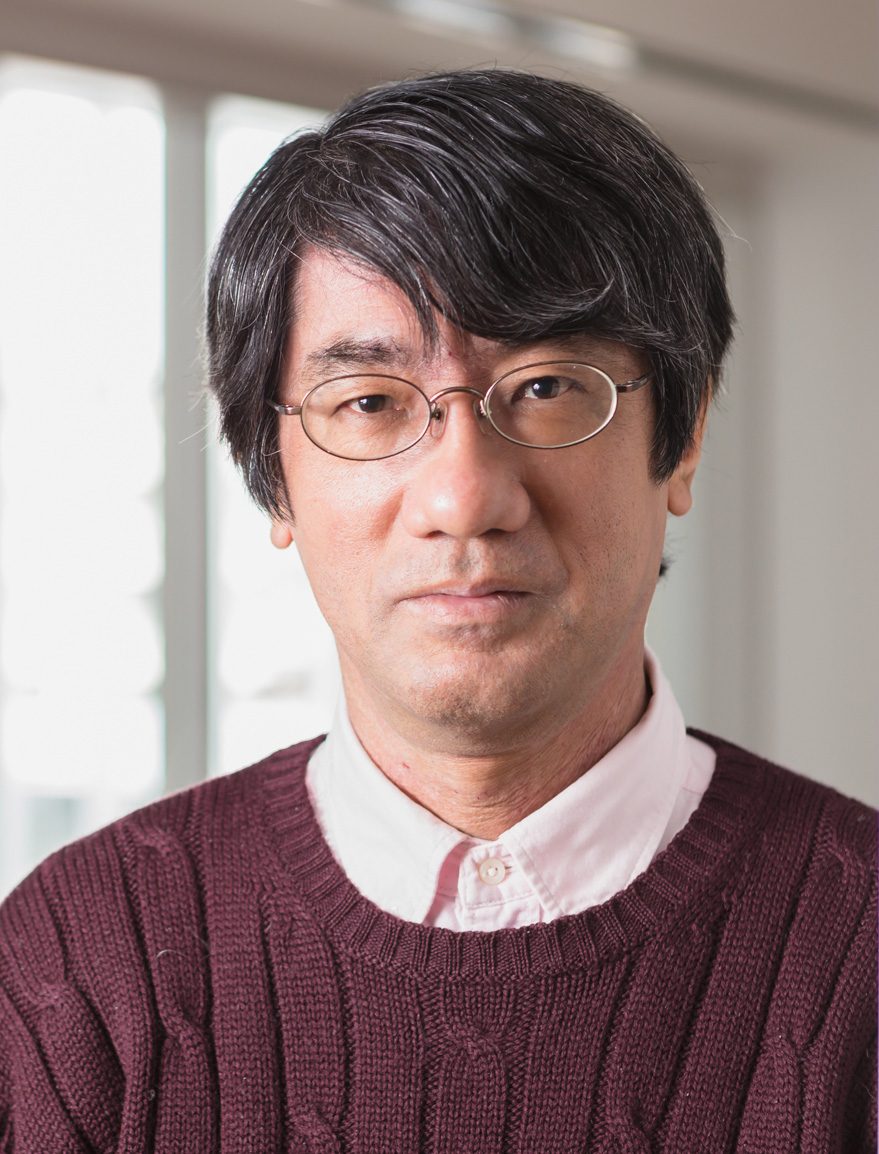}}]{Satoshi Kuriki}
received the Bachelor and Ph.D. degrees from University of Tokyo, Japan, in 1982 and 1993, respectively.  He is a Professor with the Institute of Statistical Mathematics (ISM), Tokyo, Japan, where he is also serving as Director of the Department of Mathematical Analysis and Statistical Inference.  His current major research interests include geometry of random fields, multivariate analysis, multiple comparisons, graphical models, optimal designs, and genetic statistics.
\end{IEEEbiography}

\begin{IEEEbiography}[{\includegraphics[width=1in,height=1.25in,clip,keepaspectratio]{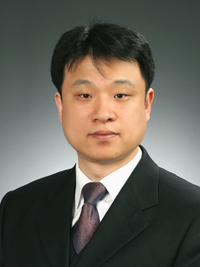}}]{Hyungdong Shin}(S'01-M'04-SM'11) received the B.S. degree in electronics engineering from Kyung Hee University, Korea, in 1999, and the M.S. and Ph.D. degrees in electrical engineering from Seoul National University, Korea, in 2001 and 2004, respectively. During his postdoctoral research at the Massachusetts Institute of Technology (MIT) from 2004 to 2006, he was with the Wireless Communication and Network Sciences Laboratory within the Laboratory for Information Decision Systems (LIDS). 
In 2006, Dr. Shin joined Kyung Hee University, Korea, where he is now an Associate Professor at the Department of Electronics and Radio Engineering. His research interests include wireless communications and information theory with current emphasis on MIMO systems, cooperative and cognitive communications, network interference, vehicular communication networks, location-aware radios and networks, physical-layer security, molecular communications.
Dr. Shin was honored with the Knowledge Creation Award in the field of Computer Science from Korean Ministry of Education, Science and Technology (2010). He received the IEEE Communications Society Guglielmo Marconi Prize Paper Award (2008) and William R. Bennett Prize Paper Award (2012). He served as a Technical Program Co-chair for the IEEE WCNC (2009 PHY Track) and the IEEE Globecom (Communication Theory Symposium, 2012). He was an Editor for IEEE Transactions on Wireless Communications (2007-2012). He is currently an Editor for IEEE Communications Letters. 
\end{IEEEbiography}

\begin{IEEEbiography}[{\includegraphics[width=1in,height=1.25in,clip,keepaspectratio]{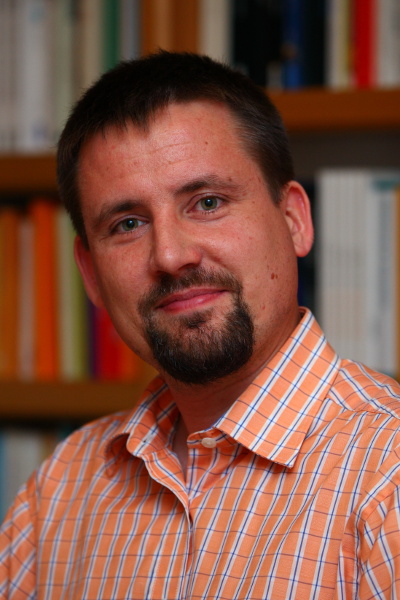}}]{Christoph Koutschan}
received the Master degree in Computer Science from
Friedrich-Alexander University in Erlangen, Germany, and the Ph.D.
degree in Symbolic Computation from the Johannes Kepler University in
Linz, Austria. He worked as a researcher at the Research Institute for
Symbolic Computation (RISC, Linz, Austria), at Tulane University (New
Orleans, USA), and at INRIA (Institut national de recherche en
informatique et en automatique, France). Currently he is with the Johann
Radon Institute for Computational and Applied Mathematics (RICAM) of the
Austrian Academy of Sciences. His research interests are on methods
related to the holonomic systems approach, particularly symbolic
summation and integration algorithms, and their application to
problems from combinatorics, knot theory, special functions, numerical
analysis, and statistical physics.
\end{IEEEbiography}

\end{document}